\documentclass[lettersize,journal]{IEEEtran}
\usepackage{amsmath,amsfonts}
\usepackage{algorithm}
\usepackage{algpseudocode}
\usepackage{array}
\usepackage[caption=false,font=normalsize,labelfont=sf,textfont=sf]{subfig}
\usepackage{textcomp}
\usepackage{stfloats}
\usepackage{url}
\usepackage{verbatim}
\usepackage{graphicx}
\usepackage{cite}
\usepackage{multirow}
\begin{document}

\title{Characterizing Statistical Separability in TP-CRIV \\for Probabilistic AI Models}

\author{Teruki Sano, 
        Minoru Kuribayashi, %~\IEEEmembership{Senior Member,~IEEE,}
        Masao Sakai,
        Shuji Isobe,
        Eisuke Koizumi,
        Zhang Zhang,
        Satoru Matsumoto
        \thanks{Teruki Sano is with Graduate School of Information Sciences, Tohoku University, Japan, and the other authors are with Center for Data-driven Science and Artificial Intelligence, Tohoku University, Japan.}
        \thanks{This study was supported by the JSPS KAKENHI (25K15225, 26H02540) and Support Center for Advanced Telecommunications Technology Research (SCAT), Japan.}%
}

% % column for its text to clear the IEEEpubid mark.

\maketitle

\begin{abstract}
Third-party challenge--response identity verification (TP-CRIV) enables an
independent verifier to assess whether a claimant possesses a model
identical to a remotely deployed model without directly accessing the
reference model.
However, for probabilistic AI models, repeated executions of the same query may produce different outputs and therefore different verification observations. This raises the question of how such stochastic evidence should be accumulated and how much evidence is required for reliable verification.

In this work, we characterize statistical separability in
TP-CRIV of probabilistic AI models.
Specifically, we relate challenge-wise behavior of matching and
non-matching provers to verification-level separability.
The characterization explicitly describes how the numbers of
independent challenges and repeated responses affect detection
performance and enables the verification budget required for a target
AUC to be estimated.
We instantiate the proposed characterization for LLMs using
open-ended challenges.
The experiments demonstrate matching--non-matching separation, close
agreement between theoretical and empirical AUCs, and consistent estimates of the minimum
verification budgets.
These results provide a statistical basis for relating probabilistic
model behavior to verification-level separability and the evidence
required for third-party verification.
\end{abstract}


\section{Introduction}
AI models are increasingly deployed through machine-learning-as-a-service
(MLaaS) platforms, where their internal parameters and implementations are
hidden from users and only input-output behavior is externally observable.
Under such access, a broad line of model verification and
fingerprinting research has investigated how model-specific behavior can be
used to identify or verify a deployed model \cite{Finger_survey}.
Existing approaches construct or identify discriminative probes, fingerprints,
or behavioral features and compare the resulting behavior with reference
information derived from a known model \cite{IPGuard,DeepJudge}.
For probabilistic generative models such as diffusion models and large language models (LLMs), repeated outputs and their resulting behavioral or feature distributions have also been explored as sources of model-specific information \cite{PromptLA,OneToken}.
In many of these settings, however, the party performing verification has
access to the reference model itself or to reference behavior obtained from it.

Third-party challenge-response identity verification (TP-CRIV)~\cite{TP-CRIV} considers a
different verification scenario.
Rather than requiring the verifier to possess the reference model, a claimant,
referred to as the prover, possesses a local model and claims that it is
identical to a model deployed by an MLaaS provider.
An independent verifier assesses this claim by issuing challenges to the prover
and evaluating the resulting evidence through the ordinary input-output
interface of the MLaaS service.
The MLaaS provider is not required to participate in the verification beyond
providing its standard inference interface.
In this way, TP-CRIV enables a model holder to demonstrate its claimed
relationship with a remotely deployed model to an independent third party.

Previous TP-CRIV studies have mainly considered approximately deterministic
verification behavior in convolutional neural network classifiers.
Extending TP-CRIV to probabilistic AI models introduces a fundamental
difficulty.
For such models, repeated executions of the same query may yield different verification-relevant outcomes, making a single challenge-response insufficient for reliable verification. Verification must therefore accumulate stochastic evidence across multiple observations.
% An observed disagreement can arise either because the prover model and the
% MLaaS model are different or simply because of stochastic variation in the
% verification-relevant behavior.
Verification of probabilistic AI models must therefore evaluate statistical
behavior accumulated over multiple observations rather than rely on an
individual response.

This requirement immediately introduces a second problem:
how much evidence is sufficient for reliable verification?
The verifier may obtain additional evidence either by issuing more independent
challenges or by repeatedly observing the behavior associated with the same
challenge.
Although both consume queries to the MLaaS service, they provide different
forms of statistical evidence.
A principled probabilistic verification framework must characterize how the
underlying matching and non-matching behaviors and the allocation of
observations jointly determine verification performance.

Rather than proposing a complete end-to-end verification framework, this work focuses on a more specific question: how to characterize statistical separability in TP-CRIV of probabilistic AI models from a finite number of stochastic responses. This work considers a setting in which the verifier specifies a target occurrence probability of a verification event for each challenge. The prover returns a witness intended to satisfy the specified probability, and the verifier observes the resulting behavior through the MLaaS input-output interface. However, because the underlying probability on the MLaaS model is latent, directly estimating its squared discrepancy from a finite sample introduces bias. To address this issue, we construct a U-statistic \cite{U-statistics} that provides an unbiased estimate of the squared probabilistic discrepancy. We use these challenge-wise statistics to define a verification score and analyze its distributions for matching and non-matching prover populations. This analysis characterizes how the numbers of independent challenges and repeated responses per challenge affect their separation and the resulting verification performance.

Furthermore, the derived statistical characterization can be used to estimate the amount of evidence required to achieve a desired detection performance. Specifically, it estimates the required numbers of challenges and responses under a given verification setting. The verification budget can therefore be determined according to a target performance rather than treated as an arbitrary experimental parameter.

To evaluate these theoretical results under a concrete and practical instantiation, we consider LLMs under stochastic generation. Each challenge consists of an open-ended prompt, a target word, and a target occurrence probability. Using its local model, the prover generates a short suffix to control the target-word occurrence probability, exploiting the model-specific probabilistic response landscape. Such control is expected to transfer more accurately for matching than for non-matching provers. Experiments with five LLMs demonstrate that the resulting matching–non-matching separability is well characterized by the proposed theory and that the required verification budgets can also be estimated accurately.

The main contributions of this work are summarized as follows.
\begin{enumerate}
    \item We formulate TP-CRIV with stochastic output as a verifier-specified occurrence-probability control problem and develop a U-statistic-based method to estimate the resulting discrepancy without finite-sample bias.
    
    \item We theoretically characterize how the challenge-wise probabilistic behavior of matching and non-matching prover populations determine verification-level detection performance.

    \item We characterize the amount and allocation of verification evidence required to achieve a desired detection performance, enabling principled verification-budget design.

   \item We instantiate the analysis for LLMs using open-ended challenges. The experiments demonstrate matching--non-matching separation, show close agreement between theoretical and empirical detection performance, and translate the observed stochastic behavior into estimates of the required verification budget.
\end{enumerate}

%%%%%%%%%%%%%%%%%%%%%%%%%%%%%%%%%%%%%
\section{Related Works}
Model identity verification has been studied under different assumptions
regarding which party possesses the reference model and how model-specific
reference information can be obtained.
These assumptions fundamentally determine what evidence can be used for
verification.
We therefore first describe the flow of TP-CRIV~\cite{TP-CRIV}, which defines the verification setting considered in this work, and then review related model fingerprinting studies~\cite{PromptLA,CollapsedGeneration,llmMap,TRAP,ESF,RESF,llmPrint,OneToken}.

\subsection{Third-Party Challenge--Response Identity Verification}
TP-CRIV~\cite{TP-CRIV} considers a setting involving a prover, an independent verifier, and an
MLaaS provider.
The prover possesses a local model and claims that it is identical to a
model deployed by the MLaaS provider.
The verifier assesses this claim without accessing the prover's model and interacts with the remotely deployed model only through the ordinary MLaaS input-output interface.
A central feature of TP-CRIV is that the verifier evaluates the prover's capability to generate model-dependent evidence in response to fresh challenges, rather than relying on a precomputed reference fingerprint. The resulting witness is then evaluated through the MLaaS interface.

For such a challenge--response mechanism to provide evidence of model
identity, the task assigned to the prover must be tied to model-specific
behavior.
In particular, it is not sufficient merely to require the prover to
produce an input that controls some output property.
The required control should be designed such that the control component
generated using the claimed model is expected to induce a corresponding
behavior when transferred to the same model deployed on the MLaaS platform, while generally producing different behavior when it is
generated using a non-matching model.
Thus, model-dependent control serves as a means of converting access to
the claimed model into fresh evidence that can subsequently be evaluated
by the verifier.

A central distinction from existing fingerprinting is that the reference model is held by the prover rather than the verifier. Whereas fingerprinting can derive model-specific reference information in advance from a known model, TP-CRIV requires the prover to generate fresh model-dependent evidence for each verifier-selected challenge.

The previous TP-CRIV formulation considered
verification-relevant behavior that can be assessed approximately
deterministically for each challenge, e.g. logits in classification models.
This assumption becomes restrictive for probabilistic generative models,
where repeated executions of the same witness may lead to different
verification observations.
We extend the TP-CRIV procedure from deterministic identity verification to the verification of probabilistic AI models.

\subsection{Model Fingerprinting for Probabilistic Models}

Fingerprinting probabilistic generative models is challenging because
repeated executions of the same input may produce different outputs.
This stochasticity appears across different model families, including
text-to-image diffusion models and LLMs, and therefore
requires verification methods that remain reliable under output variation.

For text-to-image diffusion models, PromptLA~\cite{PromptLA} performs black-box integrity
verification by comparing distributions of features extracted from
repeatedly generated images.
Other studies exploit model-specific consistency across different
random seeds as an intrinsic behavioral fingerprint of diffusion
models \cite{CollapsedGeneration}.

For LLMs, several fingerprinting approaches likewise seek model-specific
behaviors that remain reliably distinguishable despite stochastic
generation.
LLMmap~\cite{llmMap} employs active fingerprinting based on carefully selected queries
whose responses are discriminative across different model versions while
remaining relatively consistent for the same model under different
prompting configurations and stochastic sampling conditions.
TRAP~\cite{TRAP} constructs model-specific adversarial prompts that induce a predefined target response on a reference model while exhibiting low transferability to other models, thereby producing highly discriminative response patterns.
ESF~\cite{ESF} and RESF~\cite{RESF} address stochastic verification by incorporating reference-dependent response variability and statistical testing for distinguishing model modification from benign stochastic variation.
LLMPrint~\cite{llmPrint} similarly optimizes fingerprint prompts to induce consistent
model-specific token preferences that can be reliably recovered despite
output variation and model post-processing.
Although these methods differ in their mechanisms, they share the general
objective of extracting or constructing model-specific behavior that
remains sufficiently stable for reliable fingerprint verification.

In contrast, Bruckner proposed a behavioral fingerprinting approach that
directly exploits stochastic response variation itself as model-specific
information \cite{OneToken}.
Rather than seeking fingerprint queries whose outputs remain stable, the
method repeatedly submits simple open-ended prompts that admit multiple
plausible answers and characterizes each model by the empirical
distribution of its responses.
Different LLMs exhibit distinct response tendencies to such ambiguous
prompts, and the resulting empirical distributions are used as behavioral
fingerprints.
Thus, stochastic variation is not treated merely as noise to be
suppressed or tolerated, but as part of the model-specific information
used for identification.

These studies are relevant to the present work in two respects.
First, they demonstrate that externally observable LLM behavior contains
model-dependent structure, including both stable response characteristics
and model-specific stochastic response distributions.
Second, they show several ways of handling probabilistic verification
observations, ranging from allowing a set of possible responses to
distribution-level statistical inference and direct use of empirical
response distributions.

Their verification setting, however, differs from TP-CRIV.
These approaches generally rely on model-specific
reference information established in advance from a known reference
model.
For methods that construct or optimize discriminative fingerprints, such
information may consist of carefully selected queries, prompts, or
expected responses.
If this information were leaked or otherwise obtained by a malicious
user, it could potentially be reused for verification without possession
of the underlying model.
For behavioral-distribution fingerprinting, the reference information
instead takes the form of a trusted empirical response distribution,
which must be obtained from the reference model through sufficient
observations.

%%%%%%%%%%%%%%%%%%%%%%%%%%%%%%%%%%%%%%%%
\section{Verification Setting}
\label{sec:verification_setting}
\begin{figure*}[tb]
\centering
\includegraphics[width=0.8\linewidth]{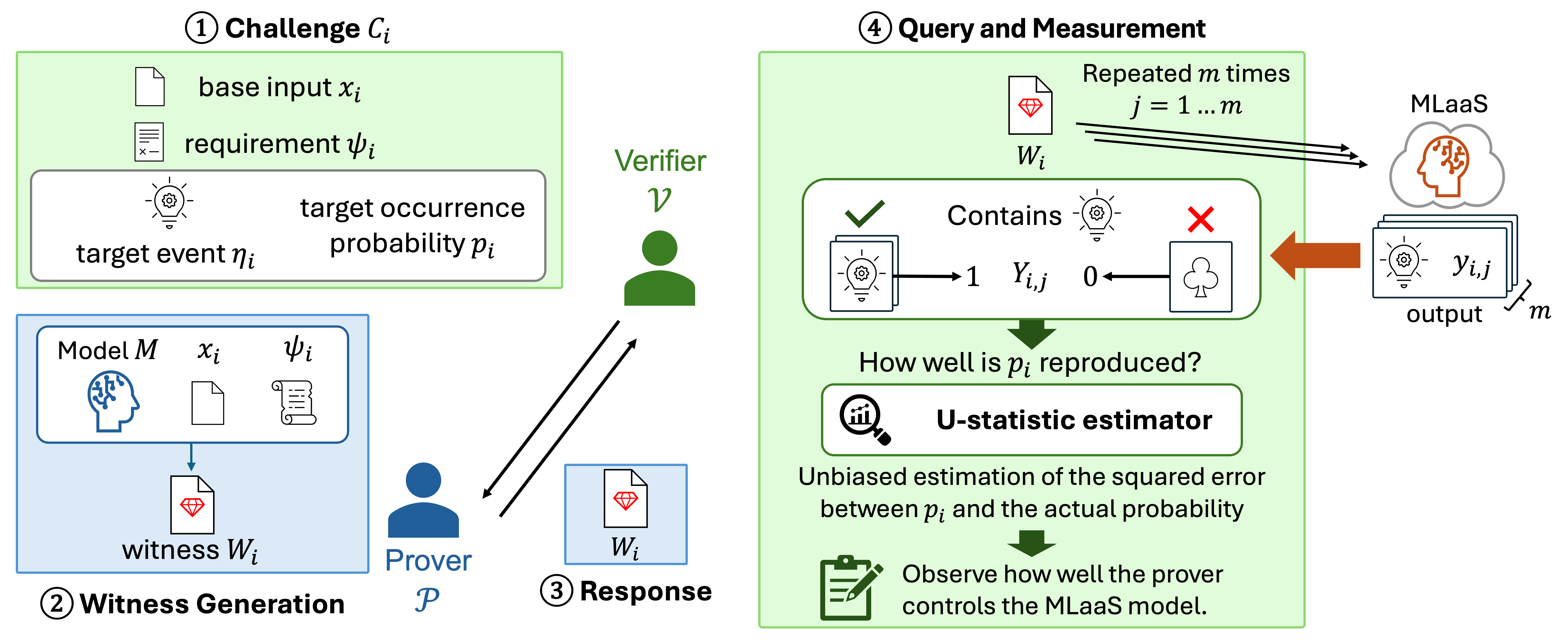}
\caption{
Overview of the setting in the TP-CRIV procedure and the proposed observation procedure for probabilistic AI models. The illustrated procedure corresponds to the $i$-th challenge.
}
\label{fig:verification_procedure}
\end{figure*}

\subsection{Setting and Assumptions}
Figure~\ref{fig:verification_procedure} illustrates the probabilistic setting in the TP-CRIV procedure and verification procedure considered in this work.
We consider a three-party verification setting consisting of a prover,
a verifier, and an MLaaS provider.

\begin{itemize}
    \item \textbf{The prover $\mathcal{P}$} possesses a model $M$ and claims that it is identical to $M^{\mathrm{MLaaS}}$ deployed by the MLaaS provider.
    Depending on the actual relationship between the two models, $\mathcal{P}$ is regarded as either matching or non-matching.
    \item \textbf{The verifier $\mathcal{V}$} aims to assess the prover's claim.
    $\mathcal{V}$ has no access to the prover's model $M$ and can access $M^{\mathrm{MLaaS}}$ only through the MLaaS input-output interface.
    \item \textbf{The MLaaS service} hosts $M^{\mathrm{MLaaS}}$ and returns an output for each query submitted by $\mathcal{V}$. It is accessible only through its ordinary inference interface and provides no verification-specific functionality.
\end{itemize}
In this work, we focus on verification settings in which a
verification-relevant event occurs probabilistically, resulting in
stochastic observations across repeated executions.
Importantly, the scope of this work is determined by the stochasticity
of the verification-relevant observation rather than by the general
stochasticity of the model output.
Specifically, the event outcome may vary across repeated executions of
the same query, and each observation is represented as a binary value
indicating whether the verification event occurs.
Even if a model generates stochastic outputs, the present analysis is
not required when the verification-relevant observation is deterministic
for each query. We instead consider situations in which repeated
observations are necessary to characterize the verification-relevant
behavior.

We further assume that repeated observations for the same query under fixed generation conditions are independent and identically distributed (i.i.d.). Accordingly, the occurrence probability of the verification event remains unchanged across repeated executions. This assumption excludes stateful or adaptive behavior in which the MLaaS model changes its response distribution based on previous queries or outputs.

\subsection{Probabilistic Challenge--Response Procedure}
\label{subsec:probabilistic_cr}
Because the verification event is observed stochastically, a single response is insufficient to characterize the prover's behavior. We therefore formulate each challenge by specifying a target occurrence probability for the verification event.
The procedure consists of the following steps.
\begin{enumerate}
    \item $\mathcal{V}$ prepares $N$ probabilistic challenges
    $C_i$, $i=1,\ldots,N$, where
    \begin{equation}
    C_i=(x_i,\psi_i).\notag
    \label{eq:challenge}
    \end{equation}
    Here, $x_i$ denotes a base input and $\psi_i$ denotes the probabilistic
    requirement imposed on $\mathcal{P}$.
    In the considered setting,
    \begin{equation}
    \psi_i=(\eta_i,p_i), \notag
    \label{eq:prob_requirement}
    \end{equation}
    where $\eta_i$ specifies an event defined on the model output and
    $p_i$ denotes its target occurrence probability.

    \item Given $C_i$, $\mathcal{P}$ applies a witness-generation procedure
    $\operatorname{ResGen}$ to its local model $M$:
    \begin{equation}
    w_i=\operatorname{ResGen}(M,C_i). \notag
    \end{equation}
    The procedure aims to generate a control component $w_i$ such that
    \begin{equation}
    \Pr[\eta_i\left(M(x_i\oplus w_i)\right)=1]\approx p_i, \notag
    \end{equation}
    where $\oplus$ denotes a model- and mechanism-specific operation that
    combines $x_i$ with the generated control component
    $w_i$. 

    \item After generating $w_i$, $\mathcal{P}$ returns a response to $\mathcal{V}$.
    As a representative form, we express the returned witness as
    \begin{equation}
        W_i=x_i\oplus w_i. \notag
    \end{equation}
    The exact response format depends on the verification mechanism.
    For example, $\mathcal{P}$ may return the complete witness $W_i$, the
    pair $(x_i,w_i)$, or only $w_i$ when $x_i$ is already available to $\mathcal{V}$. In each case, $\mathcal{V}$ obtains or reconstructs  $W_i$ to be submitted to the MLaaS service.

    \item After obtaining $W_i$, $\mathcal{V}$ submits it to the MLaaS service
    $m\geq 2$ times and receives
    \begin{equation}
    y_{i,j}=F^{\mathrm{MLaaS}}(W_i),
    \qquad j=1,\ldots,m, \notag
    \end{equation}
    where $F^{\mathrm{MLaaS}}$ denotes the service's input--output
    interface. Each output is converted into a binary observation:
    \begin{equation}
    Y_{i,j}=\eta_i(y_{i,j}), \qquad
    \eta_i(y)=
    \begin{cases}
    1, & \text{if event } i \text{ occurs},\\
    0, & \text{otherwise}.
    \end{cases}. \notag
    \label{eq:binary_observation}
    \end{equation}
    The underlying event occurrence probability is
    \begin{equation}
    q_i=\Pr[\eta_i(F^{\mathrm{MLaaS}}(W_i))=1]. \notag
    \label{eq:latent_q}
    \end{equation}
    
    Since $\mathcal{V}$ obtains only stochastic observations $Y_{i,j}$ of the verification event, rather than directly observing $q_i$, it estimates the squared discrepancy $(p_i-q_i)^2$ using the second-order U-statistic
    \begin{equation}
    U_i=
    \frac{1}{m(m-1)}
    \sum_{\substack{a\neq b}}^{m}
    (Y_{i,a}-p_i)(Y_{i,b}-p_i).
    \label{eq:challenge_u_statistic}
    \end{equation}
    Finally, $\mathcal{V}$ aggregates these observations across the $N$
    challenges to obtain the verification score:
    \begin{equation}
    T_{N,m}=\frac{1}{N}\sum_{i=1}^{N}U_i.
    \label{eq:aggregate_score}
    \end{equation}
\end{enumerate}
Given the defined matching and non-matching prover populations and
their witness-generation methods, we consider challenges under which
matching provers are expected to achieve smaller squared probabilistic
discrepancies on average than non-matching provers.

To characterize how this underlying separation translates into
verification performance, we next analyze the statistical properties
of $U_i$ and the aggregate score $T_{N,m}$ under finite observations.
This analysis relates the latent discrepancy $(p_i-q_i)^2$ to
detection performance and provides estimates of the numbers of
challenges and responses required to achieve a target performance.

%%%%%%%%%%%%%%%%%%%%%%%%%%%%%%%%%%%%%%%%
\section{Theoretical Analysis of U-Statistic-Based Verification}
\label{sec:statistical_observation}

We analyze the statistical separability of matching and
non-matching prover populations using the U-statistic-based
verification score introduced above.
For each population, we consider the challenge-wise joint
distribution of the target probability $p_i$ and the occurrence
probability $q_i$ induced on the MLaaS model by the generated witness.
Given these distributions, we derive a theoretical approximation to the area under the curve (AUC) as a function of the number of independent challenges $N$ and the number of responses per challenge $m$. This characterization is conditional on the specified challenge-wise distributions; it does not establish that a particular challenge design or witness-generation procedure necessarily produces a positive matching–non-matching error gap. 
This relationship also allows the required verification budget
to be estimated from a prescribed minimum AUC.
Specifically, we derive $N_{\min}$ for a fixed $m$ and $m_{\min}$
for a fixed $N$ under the same approximation.
To establish these results, we first derive the expectation and
variance of the challenge-wise U-statistic, then characterize
the distribution of the aggregate score and the resulting AUC.

\subsection{Statistical Properties of the Challenge-Wise U-Statistic}

For a single challenge, we omit the challenge index and treat
the target probability $p$ and the latent occurrence probability
$q$ as fixed.
Under the assumed response model,
$Y_1,\ldots,Y_m \overset{\mathrm{i.i.d.}}{\sim}\mathrm{Bernoulli}(q)$.
The U-statistic defined above can be written as
\begin{equation}
U
=
\frac{1}{m(m-1)}
\sum_{a\neq b}
(Y_a-p)(Y_b-p), \notag
\label{eq:u_statistic}
\end{equation}
where the sum is taken over all ordered pairs of distinct observations.
For $a\neq b$, independence gives
$$
\mathbb{E}[Y_aY_b]
=
\mathbb{E}[Y_a]\mathbb{E}[Y_b]
=
q^2.
$$
Therefore,
$$
\mathbb{E}[(Y_a-p)(Y_b-p)]
=
q^2-2pq+p^2
=
(p-q)^2,
$$
and hence
\begin{equation}
\mathbb{E}[U]
=
(p-q)^2.
\label{eq:u_unbiased}
\end{equation}
Therefore, for any $m\geq2$, $U$ is an unbiased estimator of the
squared probability-control error.

\subsection{Detection Based on U-Statistic Observations}
\label{subsec:u_detection}

We next characterize the separation between matching and non-matching
provers based on the challenge-wise U-statistic observations.
Suppose that $\mathcal{V}$ prepares $N$ independent challenges
$C_i$, $i=1,\ldots,N$.
For challenge $C_i$, $\mathcal{V}$ specifies a target probability
$p_i$, while $\mathcal{P}$ induces an underlying occurrence probability
$q_i$ for the designated event.
$\mathcal{V}$ queries the MLaaS model $m$ times and computes the
challenge-wise $U_i$.

The aggregated verification score $T_{N,m}$ is defined in Eq. \eqref{eq:aggregate_score}.
Since
$
\mathbb{E}[U_i]
=
(p_i-q_i)^2,
$
the expectation of the aggregated score is
$$
\mathbb{E}[T_{N,m}]
=
\frac{1}{N}
\sum_{i=1}^{N}
(p_i-q_i)^2.
$$
Thus, $T_{N,m}$ estimates the average squared probability-control
error across the verification challenges.

We denote the matching and non-matching prover populations by
$\mathtt{T}$ and $\mathtt{F}$, respectively.
For each population $X\in\{\mathtt{T},\mathtt{F}\}$, let $q_i^{X}$ denote the event
occurrence probability induced on the MLaaS model by a witness
generated by a prover from that population for challenge $C_i$.
The expected challenge-wise squared probability-control error is
defined as
\begin{equation}
\mathcal{E}^{X}
=
\mathbb{E}\left[(q_i^{X}-p_i)^2\right],
\qquad X\in\{\mathtt{T},\mathtt{F}\},
\label{eq:class_error}
\end{equation}
where the expectation is taken over the challenge-wise distribution
of $(p_i,q_i^{X})$.
The corresponding aggregate verification score $T_{N,m}^{X}$
therefore satisfies
\begin{equation}
\mathbb{E}[T_{N,m}^{X}]=\mathcal{E}^{X}. \notag
\end{equation}
To characterize the distribution of the aggregated verification
score, we next derive its variance from the challenge-wise
U-statistic observations.

For $C_i$, define the latent probability-control error as
\begin{equation}
    e_i
=
q_i-p_i,
\label{eq:e_i}
\end{equation}
and let
$
X_{i,a}
=
Y_{i,a}-p_i.
$
Then,
$
\mathbb{E}[X_{i,a}\mid p_i,q_i]
=
e_i.
$
Since $Y_{i,a}$ follows a Bernoulli distribution with parameter $q_i$,
$$
\mathrm{Var}[X_{i,a}\mid p_i,q_i]
=
q_i(1-q_i).
$$

The challenge-wise U-statistic can equivalently be written using
unordered pairs as
$$
U_i
=
\frac{1}{\binom{m}{2}}
\sum_{a<b}
X_{i,a}X_{i,b}.
$$
This expression is equivalent to the ordered-pair form because each
unordered pair appears twice in
$\sum_{a\neq b}X_{i,a}X_{i,b}$.

We define
$
s_i
=
q_i(1-q_i).
$
For distinct $a$ and $b$,
$$
\mathrm{Var}
\left[
X_{i,a}X_{i,b}
\mid p_i,q_i
\right]
=
s_i^2
+
2e_i^2s_i.
$$
Furthermore, two product terms sharing one observation are correlated.
For example,
$$
\mathrm{Cov}
\left[
X_{i,1}X_{i,2},
X_{i,1}X_{i,3}
\mid p_i,q_i
\right]
=
e_i^2s_i.
$$
In contrast, two product terms having no common observation are
independent and therefore have zero covariance.

Using the number of pair terms and overlapping pair combinations,
the conditional variance of $U_i$ is
\begin{equation}
\mathrm{Var}[U_i\mid p_i,q_i]
=
\frac{4}{m}
e_i^2q_i(1-q_i)
+
\frac{2}{m(m-1)}
q_i^2(1-q_i)^2.
\label{eq:u_conditional_variance}
\end{equation}

The above expression represents the stochastic variation of the
U-statistic for fixed $p_i$ and $q_i$.
Across challenges, however, the latent probability-control error
$e_i$, defined in Eq. \eqref{eq:e_i}, may also vary.
Applying the law of total variance gives
$$
\mathrm{Var}[U_i]
=
\mathrm{Var}
\left[
\mathbb{E}[U_i\mid p_i,q_i]
\right]
+
\mathbb{E}
\left[
\mathrm{Var}[U_i\mid p_i,q_i]
\right].
$$
Using Eqs. \eqref{eq:u_unbiased},\eqref{eq:e_i} and \eqref{eq:u_conditional_variance}, 
we obtain
\begin{gather}
\mathrm{Var}[U_i]
=
\mathrm{Var}[e_i^2]
+
\frac{4}{m}
\mathbb{E}
\left[
e_i^2q_i(1-q_i)
\right]\\ \notag
+
\frac{2}{m(m-1)}
\mathbb{E}
\left[
q_i^2(1-q_i)^2
\right].
\label{eq:u_variance}
\end{gather}
The expectations in the above expression are taken over the
challenge-wise distribution of $(p_i,q_i)$ for the considered prover
class.
The first term represents the intrinsic challenge-to-challenge
variation of the latent squared probability-control error, whereas
the latter two terms represent the stochastic variation caused by
finite response sampling.

For prover class $X\in\{\mathtt{T},\mathtt{F}\}$, define
$W^{X}(m)
=
\mathrm{Var}[U_i^{X}].\notag
\label{eq:w_definition}$
More explicitly,
\begin{gather}
\label{eq:w_explicit}
W^{X}(m)
=
\mathrm{Var}
\left[
(e_i^{X})^2
\right]
+
\frac{4}{m}
\mathbb{E}
\left[
(e_i^{X})^2
q_i^{X}(1-q_i^{X})
\right]
\\ \notag +
\frac{2}{m(m-1)}
\mathbb{E}
\left[
(q_i^{X})^2
(1-q_i^{X})^2
\right].\notag
\end{gather}
Because the aggregated verification score $T_{N,m}^{X}$ is given by Eq.\eqref{eq:aggregate_score}
and the challenge-wise observations are assumed to be independent and
drawn from the same prover-class distribution,
\begin{equation}
\mathrm{Var}[T_{N,m}^{X}]
=
\frac{W^{X}(m)}{N}. \notag
\label{eq:aggregate_variance}
\end{equation}

Under these assumptions and for sufficiently large $N$, the central
limit theorem gives
\begin{equation}
T_{N,m}^{X}
\approx
\mathcal{N}
\left(
\mathcal{E}^{X},
\frac{W^{X}(m)}{N}
\right),\notag
\label{eq:aggregate_gaussian}
\end{equation}
where $\mathcal{N}(\mu,\sigma^2)$ denotes a normal distribution with
mean $\mu$ and variance $\sigma^2$.

Because a smaller verification score indicates better agreement with
the verifier-specified probabilities, the AUC
can be expressed as
$
\mathrm{AUC}
=
\Pr
\left[
T_{N,m}^{\mathtt{T}}
<
T_{N,m}^{\mathtt{F}}
\right].
$
Since the difference between two independent Gaussian random
variables is also Gaussian, the AUC can be approximated as
\begin{equation}
\mathrm{AUC}(N,m)
\approx
\Phi
\left(
\frac{
\sqrt{N}
\left(
\mathcal{E}^{\mathtt{F}}
-
\mathcal{E}^{\mathtt{T}}
\right)
}{
\sqrt{
W^{\mathtt{F}}(m)
+
W^{\mathtt{T}}(m)
}
}
\right),
\label{eq:theoretical_auc}
\end{equation}
where $\Phi(\cdot)$ denotes the cumulative distribution function of
the standard normal distribution.

This expression directly relates the expected detection performance
to the underlying challenge-wise probability-control behavior of the
matching and non-matching prover classes.
The numerator represents the difference in their expected squared
probability-control errors, while the denominator accounts for both
inter-challenge variation and finite-$m$ stochastic response
variation.
Consequently, once the challenge-wise distributions of $p_i$ and
$q_i$ are characterized, the expected AUC for any given $N$ and $m$
can be theoretically estimated.

\subsection{Required Verification Budget for a Target AUC}
\label{subsubsec:required_budget_theory}

The derived AUC expression can also be used to estimate the
verification budget required to achieve a target detection
performance.
The following budget derivations assume $\mathcal{E}^{\mathtt{F}}>\mathcal{E}^{\mathtt{T}}$, meaning that matching provers achieve a smaller expected squared probability-control error than non-matching provers. This condition specifies the direction of separation required for an AUC greater than $0.5$ under the Gaussian approximation.
Let the target AUC be $\alpha$, where $0.5<\alpha<1$.
The condition
$
\mathrm{AUC}(N,m)
>
\alpha
$
is approximately equivalent to
$$
\frac{
\sqrt{N}
\left(
\mathcal{E}^{\mathtt{F}}
-
\mathcal{E}^{\mathtt{T}}
\right)
}{
\sqrt{
W^{\mathtt{F}}(m)
+
W^{\mathtt{T}}(m)
}
}
>
\Phi^{-1}(\alpha).
$$
Therefore,
$$
N
>
\frac{
\left(\Phi^{-1}(\alpha)\right)^2
\left(
W^{\mathtt{F}}(m)
+
W^{\mathtt{T}}(m)
\right)
}{
\left(
\mathcal{E}^{\mathtt{F}}
-
\mathcal{E}^{\mathtt{T}}
\right)^2
}.
$$
Since $N$ must be an integer, the minimum number of challenges
required for a fixed $m$ is
\begin{equation}
N_{\min}(m;\alpha)
=
\left\lfloor
\frac{
\left(\Phi^{-1}(\alpha)\right)^2
\left(
W^{\mathtt{F}}(m)
+
W^{\mathtt{T}}(m)
\right)
}{
\left(
\mathcal{E}^{\mathtt{F}}
-
\mathcal{E}^{\mathtt{T}}
\right)^2
}
\right\rfloor
+
1.
\label{eq:nmin_general}
\end{equation}
where $\lfloor \cdot \rfloor$ denotes the floor function.

To make the dependence on $m$ explicit, define
\begin{equation}
A^X
=
\mathrm{Var}
\left[
(e_i^{X})^2
\right],
\label{eq:A_X}
\end{equation}
\begin{equation}
B^X
=
4
\mathbb{E}
\left[
(e_i^{X})^2
q_i^{X}
(1-q_i^{X})
\right],
\label{eq:B_X}
\end{equation}
\begin{equation}
C^X
=
2
\mathbb{E}
\left[
(q_i^{X})^2
(1-q_i^{X})^2
\right].
\label{eq:C_X}
\end{equation}
Then,
\begin{equation}
W^{X}(m)
=
A^X
+
\frac{B^X}{m}
+
\frac{C^X}{m(m-1)}.
\label{eq:w_abc}
\end{equation}
Let
$$
A=A^\mathtt{T}+A^\mathtt{F},
\qquad
B=B^\mathtt{T}+B^\mathtt{F},
\qquad
C=C^\mathtt{T}+C^\mathtt{F}.
$$
Accordingly,
$$
W^{\mathtt{F}}(m)+W^{\mathtt{T}}(m)
=
A
+
\frac{B}{m}
+
\frac{C}{m(m-1)}.
$$
Substituting this expression into Eq.~\eqref{eq:nmin_general} gives
\begin{equation}
N_{\min}(m,\alpha)
=
\left\lfloor
\frac{
(\Phi^{-1}(\alpha))^2
}{
\left(
\mathcal{E}^{\mathtt{F}}
-
\mathcal{E}^{\mathtt{T}}
\right)^2
}
\left(
A
+
\frac{B}{m}
+
\frac{C}{m(m-1)}
\right)
\right\rfloor
+
1.
\label{eq:nmin_explicit}
\end{equation}
This expression explicitly describes the trade-off between the number
of challenges $N$ and the number of responses per challenge $m$.
Increasing $m$ reduces the finite-sampling terms $B/m$ and
$C/m(m-1)$, whereas the challenge-level variation $A$ remains.
% In particular,
% $$
% \lim_{m\rightarrow\infty}
% \left[
% W^{F}(m)+W^{T}(m)
% \right]
% =
% A.
% $$
Therefore, increasing the number of responses per challenge $m$ cannot
eliminate the intrinsic challenge-to-challenge variation.
This explains why the benefit of increasing $m$ eventually
diminishes.

Conversely, for a fixed $N$, the minimum number of responses per
challenge can also be obtained.
Define
\begin{equation}
R
=
\frac{
N
\left(
\mathcal{E}^{\mathtt{F}}
-
\mathcal{E}^{\mathtt{T}}
\right)^2
}{
(\Phi^{-1}(\alpha))^2
}.
\label{eq:R_definition}
\end{equation}
Then the required condition becomes
\begin{equation}
    A
+
\frac{B}{m}
+
\frac{C}{m(m-1)}
<
R.
\label{eq:form_m}
\end{equation}
Since $B\geq0$ and $C\geq0$, the finite-sampling contribution
$$
\frac{B}{m}
+
\frac{C}{m(m-1)}
$$
is non-increasing with respect to $m$ for $m\geq2$.
% More specifically,
% $$
% \left[
% \frac{B}{m+1}
% +
% \frac{C}{m(m+1)}
% \right]
% -
% \left[
% \frac{B}{m}
% +
% \frac{C}{m(m-1)}
% \right]
% =
% -
% \frac{B}{m(m+1)}
% -
% \frac{2C}{m(m-1)(m+1)}
% \leq 0.
% $$
% The inequality is strict unless $B=C=0$.
Therefore,
$$
A
+
\frac{B}{m}
+
\frac{C}{m(m-1)}
$$
decreases monotonically toward $A$ as $m$ increases. In particular, as $m\to\infty$, the above expression converges to $A$.
Consequently, a finite $m$ satisfying Eq.\eqref{eq:form_m}
exists if and only if
$
A<R.
$
Equivalently,
$$
N
>
\frac{
(\Phi^{-1}(\alpha))^2 A
}{
\left(
\mathcal{E}^{\mathtt{F}}
-
\mathcal{E}^{\mathtt{T}}
\right)^2
}.
$$
If this condition is not satisfied, increasing $m$ cannot achieve the
target AUC, because the total variance term is bounded below by $A$.
Conversely, when $R>A$, the finite-sampling terms decrease toward zero,
so a sufficiently large finite $m$ necessarily satisfies the required
inequality.

When $R>A$, define
$
D
=
R-A.
$
Then
$$
\frac{B}{m}
+
\frac{C}{m(m-1)}
<
D.
$$
Rearranging gives
$$
Dm^2
-
(D+B)m
+
(B-C)
>
0.
$$
The corresponding boundary is given by the larger root,
\begin{equation}
m_{\mathrm{bd}}(N,\alpha)
=
\frac{
D+B
+
\sqrt{
(D+B)^2
-
4D(B-C)
}
}{
2D
}.\notag
\label{eq:m_boundary}
\end{equation}
Therefore, the minimum number of responses is
\begin{equation}
m_{\min}(N,\alpha)
=
\max\left\{
2,\,
\left\lfloor m_{\mathrm{bd}}\right\rfloor+1
\right\}.
\label{eq:m_min}
\end{equation}

These expressions allow $\mathcal{V}$ to estimate the required
verification budget directly from the challenge-wise probability
behavior of matching and non-matching provers.
In particular, empirical observations of $p_i$ and $q_i$ can be used
to estimate
$\mathcal{E}^{\mathtt{T}}$,
$\mathcal{E}^{\mathtt{F}}$,
$A$,
$B$,
and $C$,
after which the theoretical AUC and the minimum required $N$ and $m$
can be calculated for a desired target AUC.

%%%%%%%%%%%%%%%%%%%%%%%%%%%%%%%%%%%%%%%%

\section{LLM Instantiation}
\label{sec:llm_instantiation}

In this section, we investigate whether the proposed TP-CRIV framework for probabilistic AI models can be instantiated for LLMs to distinguish matching from non-matching provers.
Our objective in this study is not to construct a complete operational
verification system with a calibrated decision threshold.
Instead, we focus on two questions:
whether model-dependent probabilistic challenges can produce
statistical separation between matching and non-matching provers, and
whether the resulting separability can be characterized using the
theoretical analysis developed in the previous section.

\subsection{LLM Setting}
\begin{figure*}[tb]
\centering
\includegraphics[width=0.8\linewidth]{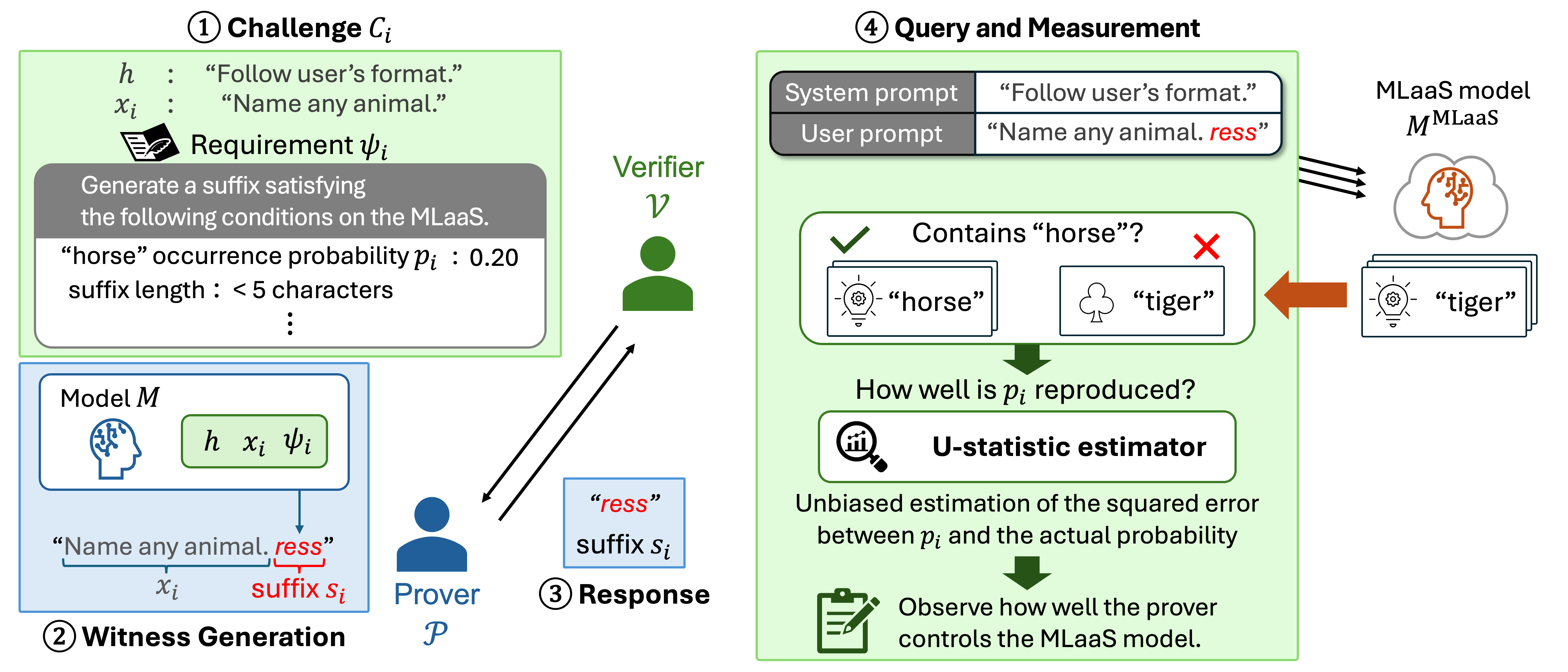}
\caption{
Overview of the LLM instantiation of TP-CRIV for probabilistic models.
}
\label{fig:llm_framework}
\end{figure*}
Figure \ref{fig:llm_framework} illustrates the overall procedure of the LLM-based instantiation considered in this work.
Both $M$ and
$M^{\mathrm{MLaaS}}$ are assumed to be LLMs operated under probabilistic
generation with a nonzero temperature.
Thus, repeated executions of the same input may produce different
outputs.

In this feasibility study, model identity is instantiated in a strict
form.
A matching prover possesses the same LLM as
$M^{\mathrm{MLaaS}}$, whereas a non-matching prover possesses a
different LLM.
We further assume that both matching and non-matching provers employ
the same suffix-generation strategy $\operatorname{ResGen}$.
That is, each prover applies the same algorithm to its own local model
under the same challenge specification.
This allows us to examine whether the resulting difference in
probabilistic behavior is attributable to the difference between the
underlying models rather than to different prover strategies.

We represent the input-output operation of a model $M$ as
$
M\!\left(\mathcal{T}_M(h,x)\right),
$
where $h$ and $x$ denote the system and user prompts, respectively,
and $\mathcal{T}_M$ is a predefined model-specific serialization
function that converts them into the input format expected by $M$.
The local model $M$ and the deployed model $M^{\mathrm{MLaaS}}$
use fixed serialization functions $\mathcal{T}_M$ and
$\mathcal{T}_{M^{\mathrm{MLaaS}}}$, respectively, specified according
to their prompt formats.
We assume that each model uses a fixed serialization function
appropriate to its prompt format, and that identical models use
the same serialization function.
These functions are specified in advance and are neither selected
nor optimized by $\mathcal{P}$.

\subsection{TP-CRIV Procedure for LLMs}
\label{subsec:llm_cr}
We instantiate the probabilistic challenge--response procedure
described in the previous section as follows.
\begin{enumerate}
    \item $\mathcal{V}$ constructs the $i$-th challenge as
$$
C_i=(h,x_i,\psi_i),
\qquad
\psi_i=(\omega_i,p_i,N_s,\mathcal{F}_i),
$$
where $h$ is a common system instruction, $x_i$ is an open-ended
user prompt, and $\psi_i$ specifies the target word $\omega_i$,
the target occurrence probability $p_i$, the maximum suffix length
$N_s$, and a set of prohibited target-related fragments $\mathcal{F}_i$.
The requirement specified by $\psi_i$ is to generate a suffix $s_i$
such that, when $s_i$ is appended to $x_i$, the response from the
MLaaS model contains $\omega_i$ with probability close to $p_i$.
The suffix must satisfy
$$
|s_i|_{\mathrm{char}}\leq N_s,
$$
and must not contain $\omega_i$ or any fragment in $\mathcal{F}_i$.
These constraints rule out directly including the target word
or the specified target-related fragments in the suffix.

\item Given $C_i$, $\mathcal{P}$ uses a suffix-generation
algorithm $\operatorname{SufGen}$, an instantiation of
$\operatorname{ResGen}$, with $M$ to obtain
$$
s_i=\operatorname{SufGen}(M,C_i).
$$
The algorithm seeks a suffix $s_i$ that satisfies the generation
constraints specified in $\psi_i$ and makes the target-word
occurrence probability under $M$ close to $p_i$:
$$
\Pr[
\omega_i \text{ appears in }
M\!\left(\mathcal{T}_M(h,x_i\Vert s_i)\right)]
\approx p_i,
$$
where the probability is taken over the model's stochastic
response generation.

\item $\mathcal{P}$ returns the generated $s_i$ to $\mathcal{V}$
as its response to $C_i$.

\item Upon receiving $s_i$, $\mathcal{V}$ checks whether it satisfies
the suffix constraints specified in $\psi_i$ and rejects the
response if any constraint is violated.
Otherwise, $\mathcal{V}$ appends $s_i$ to $x_i$ and submits
the resulting user input $x_i\Vert s_i$, where $\Vert$ denotes string concatenation, together with the common
system instruction $h$, to the MLaaS service $m$ times to obtain
output messages
$$
y_{i,j}
=
M^{\mathrm{MLaaS}}\!\left(
\mathcal{T}_{M^{\mathrm{MLaaS}}}(h,x_i\Vert s_i)
\right),
\qquad j=1,\ldots,m,
$$
where each invocation uses independent randomness for response generation.

For each response, $\mathcal{V}$ records whether the target word
$\omega_i$ appears:
$$
Y_{i,j}
=
\begin{cases}
1, & \text{if } \omega_i \text{ appears in } y_{i,j},\\
0, & \text{otherwise}.
\end{cases}
$$
Using these observations, $\mathcal{V}$ computes the challenge-wise
score $U_i$ via the U-statistic defined in Eq.~\eqref{eq:challenge_u_statistic}.

$\mathcal{V}$ repeats this procedure for $N$ challenges,
$i=1,\ldots,N$, and aggregates the resulting scores $U_i$ into
$T_{N,m}$ as defined in Eq.~\eqref{eq:aggregate_score}.
% This aggregate score quantifies the prover's ability to control
% the target-word occurrence probabilities under the MLaaS model
% toward the specified targets.

\end{enumerate}

In this feasibility study, we do not calibrate an operational decision
threshold for accepting or rejecting $\mathcal{P}$.
Instead, we evaluate whether the distributions of $T_{N,m}$ obtained
from matching and non-matching provers are statistically separable.
We further compare this empirical separability with the theoretical
AUC and the required numbers of challenges and responses derived in
the previous section.

\subsection{Challenge Design Principle}

The challenge design is motivated by the observation that LLMs exhibit
model-specific response distributions for simple open-ended prompts.
For example, a prompt such as ``Name an animal'' allows multiple
plausible responses, such as ``dog,'' ``cat,'' or ``horse,'' rather
than a single correct answer.
Previous work has shown that repeatedly querying LLMs with such
prompts can reveal characteristic output distributions that can serve
as behavioral fingerprints~\cite{OneToken}.
We exploit this property from a different perspective.
An open-ended prompt admits multiple plausible responses, and the
relative probabilities assigned to these responses depend on the
underlying model.
We refer to this model-dependent structure as a probabilistic response
landscape.
Our challenge asks the prover to locally manipulate this response
landscape.
Given an open-ended base prompt $x_i$, the prover searches for a short
suffix $s_i$ that shifts the occurrence probability of a designated
target word $\omega_i$ toward the verifier-specified value $p_i$ on
its own model.

The key hypothesis is that the transferability of this local
probability manipulation depends on the underlying model.
When $M$ and $M^{\mathrm{MLaaS}}$ are the same model, a suffix
constructed on $M$ is expected to preserve a similar probability
control effect on $M^{\mathrm{MLaaS}}$, resulting in
$q_i$ close to $p_i$.
When the prover possesses a different model, the same
suffix-generation algorithm operates on a different probabilistic
response landscape.
The resulting suffix may therefore induce a different target-word
probability after being transferred to $M^{\mathrm{MLaaS}}$.
Across multiple challenges, such differences may produce statistically
distinct score distributions for matching and non-matching provers.

The suffix constraints are intended to strengthen this
model-dependent effect.
Prohibiting the target word and its related fragments suppresses
direct lexical insertion, while restricting the suffix length limits
the amount of explicit instruction that can be embedded in the suffix.
The challenge is therefore designed to exploit subtle
model-dependent response tendencies rather than a generally
transferable explicit instruction.

Accordingly, the purpose of the proposed challenge is not to guarantee
model identification for every possible pair of LLMs.
Rather, this feasibility study investigates whether such constrained
probability-control challenges can induce measurable separation
between matching and non-matching models and whether the degree of
separation can be theoretically characterized.

\subsection{Suffix Generation Method}
\label{subsec:suffix_generation}

As a concrete instantiation of the prover's response-generation
strategy, we introduce a GCG-based suffix generator~\cite{GCG}.
Given $C_i$, the generator searches for a suffix $s_i$
that brings the occurrence probability of the target word $\omega_i$
close to $p_i$ on the prover's local model $M$, while satisfying
the constraints specified in $\psi_i$.
In particular, the search aims to achieve this probability control
under a tight character-length constraint $|s_i|_{\mathrm{char}}\leq N_s$. Algorithm~\ref{alg:prob_suffix} summarizes the overall suffix-generation procedure.

To search effectively under this constraint, we combine
gradient-guided candidate search with empirical evaluation through
stochastic response generation.
The gradient-guided stage uses a differentiable surrogate to identify
promising suffix candidates, while the empirical stage evaluates
their actual target-word occurrence probabilities and selects a
suffix whose estimated probability is closest to $p_i$.
Directly optimizing the empirical occurrence probability of the
target word is difficult because stochastic text generation is
non-differentiable.
We therefore use a differentiable teacher-forcing probability as a
surrogate during candidate search.

The original GCG method searches for adversarial suffixes by optimizing
the likelihood of a desired target completion.
In our implementation, we instead use the teacher-forcing probability
of the first token of the target completion as a differentiable
surrogate.
Specifically, let $\pi_i(s)$ denote the probability assigned by the
prover model to the first token of the target completion $\omega_i$
given the serialized input $\mathcal{T}_M(h,x_i,s)$.
The search objective is
$
L_i(s)
=
\left(
\pi_i(s)-p_i
\right)^2.
$
This surrogate does not directly represent the probability that
$\omega_i$ appears in a stochastically generated response.
Rather, the search is intended to bias the model's local output
distribution toward a state in which the overall occurrence tendency
of the target word is affected.
GCG is used to efficiently identify suffix candidates that reduce
$L_i(s)$.
The resulting candidates are then evaluated through actual stochastic
generation, and are empirically reranked according to the observed
target-word occurrence probability.

\begin{algorithm}[t]
\caption{Probability-Targeted Suffix Generation Method}
\label{alg:prob_suffix}
\begin{algorithmic}[1]
\Require Challenge $C_i=(h,x_i,\psi_i)$, prover model $M$
\Ensure Suffix $s_i$

\State Extract $\omega_i$, $p_i$, and suffix constraints from $\psi_i$

\State Generate candidate suffixes $\mathcal{S}_i$ using a GCG-based
search

\State Remove candidates that violate the suffix constraints

\State Select a shortlist $\mathcal{S}_i^{\mathrm{short}}$
according to the GCG-based surrogate objective

\ForAll{$s \in \mathcal{S}_i^{\mathrm{short}}$}
    \For{$k=1,\ldots,K_1$}
        \State
        $y_k \sim M\!\left(\mathcal{T}_M(h,x_i\Vert s)\right)$
    \EndFor
    \State
    $\displaystyle
    \hat{p}_i^{(1)}(s)
    =
    \frac{1}{K_1}
    \sum_{k=1}^{K_1}
    \mathbf{1}[\omega_i \in y_k]
    $
\EndFor

\State Retain candidates with the smallest
$|\hat{p}_i^{(1)}(s)-p_i|$

\ForAll{retained suffixes $s$}
    \For{$k=K_1+1,\ldots,K_1+K_2$}
        \State
        $y_k \sim M\!\left(\mathcal{T}_M(h,x_i\Vert s)\right)$
    \EndFor
    \State
    $\displaystyle
    \hat{p}_i(s)
    =
    \frac{1}{K_1+K_2}
    \sum_{k=1}^{K_1+K_2}
    \mathbf{1}[\omega_i \in y_k]
    $
\EndFor

\State
$\displaystyle
s_i
=
\arg\min_s
\left|
\hat{p}_i(s)-p_i
\right|
$

\State \Return $s_i$

\end{algorithmic}
\end{algorithm}

%%%%%%%%%%%%%%%%%%%%%%%%%%%%%%%%%%%%%%%%
\section{Experimental Evaluation}
\label{sec:experiments}
This section evaluates the LLM instantiation from four perspectives.
Section~\ref{subsec:probability_control_behavior} examines the
probability-control behavior of suffixes generated by matching
and non-matching provers.
Section~\ref{subsec:empirical_separation} evaluates the separation
between their verification scores using empirical AUC.
Section~\ref{subsec:theory_empirical_auc} compares the theoretical
and empirical AUCs to assess the accuracy of the approximation.
Finally, Section~\ref{subsec:required_budget_experiment} examines
the accuracy of the estimated minimum numbers of challenges and
responses required to achieve a specified target AUC.
\subsection{Experimental Setup}
\label{subsec:experimental_setup}

We evaluate the proposed challenge using five LLMs:
Vicuna-7B-v1.5 \cite{vicuna},
Mistral-7B-Instruct-v0.2 \cite{mistral},
Llama-2-7B-chat-hf \cite{llama},
Qwen2.5-7B-Instruct \cite{qwen}, and
Yi-6B-Chat \cite{yi}.
Hereafter, we refer to these models as
Vicuna, Mistral, Llama-2, Qwen2.5, and Yi, respectively.
Each of the five models is used both as $M$
and $M^{\mathrm{MLaaS}}$, and all combinations
are evaluated.
For each fixed $M^{\mathrm{MLaaS}}$, the same model serves as the
matching prover, while the remaining four models serve as
non-matching provers.
All five prover models employ the same suffix-generation procedure
described in the previous section.
The common generation and suffix-generation settings are summarized
in Table~\ref{tab:llm_settings} in Appendix~\ref{app:llm_settings}.

We construct a fixed set of $500$ challenges.
Each base prompt $x_i$ requests a single English instance belonging to
a semantic category without uniquely specifying which valid instance
should be returned.
The challenge set covers a diverse range of semantic categories and
uses multiple paraphrased prompt forms for each category.
For each prompt, the target word $\omega_i$ is selected from plausible
responses belonging to the corresponding category.
For each challenge, the target probability $p_i$ is sampled from
$
\{0.2,0.4,0.6,0.8\},
$
and the resulting challenge specification
$(x_i,\omega_i,p_i)$ is fixed and shared across all model combinations.
Thus, differences between matching and non-matching conditions are
evaluated using the same set of probabilistic requirements.
$\mathcal{V}$ additionally specifies the common system instruction $h$ as
\textit{``You follow the user's output-format instructions exactly.''},
which is provided to $\mathcal{P}$ before challenge execution and remains
fixed across all challenges and models.
Each LLM $M$ is associated with a predefined model-specific prompt
serialization function $\mathcal{T}_M$, whose template is provided
in Appendix~\ref{app:prompt_templates}.
Importantly, neither the system instruction $h$ nor the model-specific
template $\mathcal{T}_M$ is optimized by the $\mathcal{P}$.
Only the suffix $s_i$ is generated by the prover.

\subsection{Probability-Control Behavior}
\label{subsec:probability_control_behavior}

We first examine how the model used for suffix generation affects the
probability-control behavior on a fixed evaluation target.
For each target model, suffixes are independently generated using all
five evaluated models and
are then evaluated on the same fixed target model.
In the verification setting, the suffix-generation model corresponds
to $M$, whereas the evaluation target corresponds
to $M^{\mathrm{MLaaS}}$.
Thus, the same-model combination represents the matching condition,
while the remaining four combinations represent non-matching
conditions.
For each challenge, let $\hat{q}_i$ denote the empirical occurrence
probability of the target word on the evaluation target.
We evaluate the probability-control behavior using the squared error
$(p_i-\hat{q}_i)^2$.
For quantitative comparison across all five target models, we compute
$$
\widehat{\mathcal{E}}
=
\frac{1}{500}
\sum_{i=1}^{500}
(p_i-\hat{q}_i)^2,
\qquad
\widehat{A}
=
\mathrm{Var}\left[(p_i-\hat{q}_i)^2\right].
$$

The resulting probability-control performance is summarized in
Table~\ref{tab:probability_control_behavior} for all combinations of
the five prover models and five target models.
\begin{table}[t]
\centering
\caption{
Empirical probability-control errors for each generation and evaluation model.
Values are $\widehat{\mathcal{E}}$ with $\widehat{A}$ in parentheses.
}
\label{tab:probability_control_behavior}

\renewcommand{\arraystretch}{1.1}

\begin{tabular}{l|ccccc}
\hline
&
\multicolumn{5}{c}{\textbf{Generation model}}
\\
\cline{2-6}
\shortstack{\textbf{Evaluation}\\\textbf{model}}
&
Vicuna
&
Mistral
&
Llama-2
&
Qwen2.5
&
Yi
\\
\hline

\multirow{2}{*}{Vicuna}
&
\textbf{0.0392}
&
0.2074
&
0.2110
&
0.2203
&
0.2010
\\
&
\textbf{(0.0083)}
&
(0.0402)
&
(0.0425)
&
(0.0431)
&
(0.0410)
\\[2pt]

\multirow{2}{*}{Mistral}
&
0.2110
&
\textbf{0.0541}
&
0.2295
&
0.2480
&
0.2250
\\
&
(0.0441)
&
\textbf{(0.0173)}
&
(0.0474)
&
(0.0482)
&
(0.0446)
\\[2pt]

\multirow{2}{*}{Llama-2}
&
0.1536
&
0.2180
&
\textbf{0.1674}
&
0.2207
&
0.2249
\\
&
(0.0318)
&
(0.0453)
&
\textbf{(0.0431)}
&
(0.0462)
&
(0.0460)
\\[2pt]

\multirow{2}{*}{Qwen}
&
0.2119
&
0.2307
&
0.2508
&
\textbf{0.0420}
&
0.2373
\\
&
(0.0455)
&
(0.0466)
&
(0.0488)
&
\textbf{(0.0137)}
&
(0.0491)
\\[2pt]

\multirow{2}{*}{Yi}
&
0.1579
&
0.2212
&
0.2339
&
0.2175
&
\textbf{0.1411}
\\
&
(0.0323)
&
(0.0437)
&
(0.0449)
&
(0.0454)
&
\textbf{(0.0377)}
\\

\hline
\end{tabular}
\end{table}
The diagonal entries correspond to matching conditions and the four
off-diagonal entries to non-matching conditions.
The magnitude of the matching--non-matching difference varies across
target models, showing that the suffix-generation procedure used in
this study produces model-dependent probability-control behavior.
In particular, the matching conditions for Vicuna, Mistral, and Qwen
achieve relatively small mean squared errors respectively.
In contrast, probability control is comparatively more difficult for Llama-2 and Yi under the present suffix-generation procedure, resulting in less pronounced matching--non-matching separation than for the other evaluated models.
This procedure is one concrete prover strategy rather than a
universally reliable challenge solver; the next subsection examines
how the observed challenge-wise differences translate into
verification-level separation.

All subsequent experiments use the suffixes generated in this
experiment.
Each suffix satisfies all constraints specified in $\psi_i$,
including the character-length limit $|s_i|_{\mathrm{char}}\leq N_s$
and the exclusion of the target word and prohibited target-related
fragments.
Thus, all suffixes pass the verifier's validity check.
\subsection{Empirical Matching and Non-Matching Separation}
\label{subsec:empirical_separation}
We next investigate whether the challenge-wise probability-control
differences translate into verification-level separation.
For each fixed MLaaS model, the matching population consists of the
$500$ challenges generated using the same prover model.
For the non-matching class, all
$(\text{non-matching prover model},\text{challenge})$
instances are combined into a single pooled population.
For each Monte Carlo trial, $N$ challenge instances are sampled from
the corresponding population.
For every selected instance, $m$ stochastic responses are sampled,
the challenge-wise U-statistic is computed, and the resulting values
are averaged to obtain $T_{N,m}$.
This Monte Carlo trial is repeated $2000$ times for each
$(N,m)$ setting to construct the empirical score distributions used
for AUC evaluation.
We evaluate
$N\in\{1,3,5,10,20,50,100\}$
and
$m\in\{2,3,5,10,20,50,100\}.$
\begin{figure}[t]
\centering
\begin{minipage}[b]{0.90\columnwidth}
    \centering
    \includegraphics[width=\linewidth]{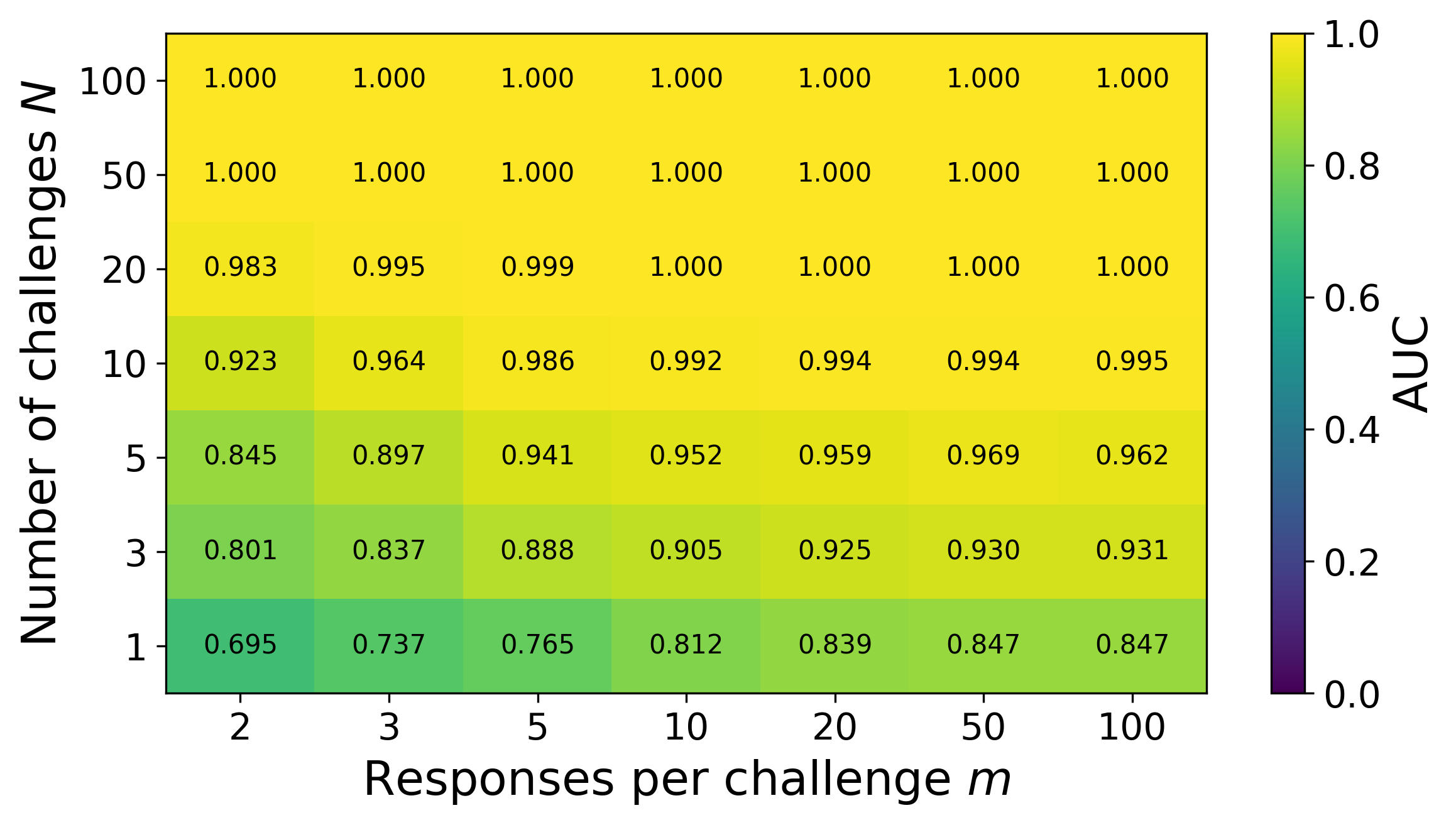}
    \smallskip
    \text{(a) $M^\textrm{MLaaS}$: Vicuna}
\end{minipage}
\begin{minipage}[b]{0.90\columnwidth}
    \centering
    \includegraphics[width=\linewidth]{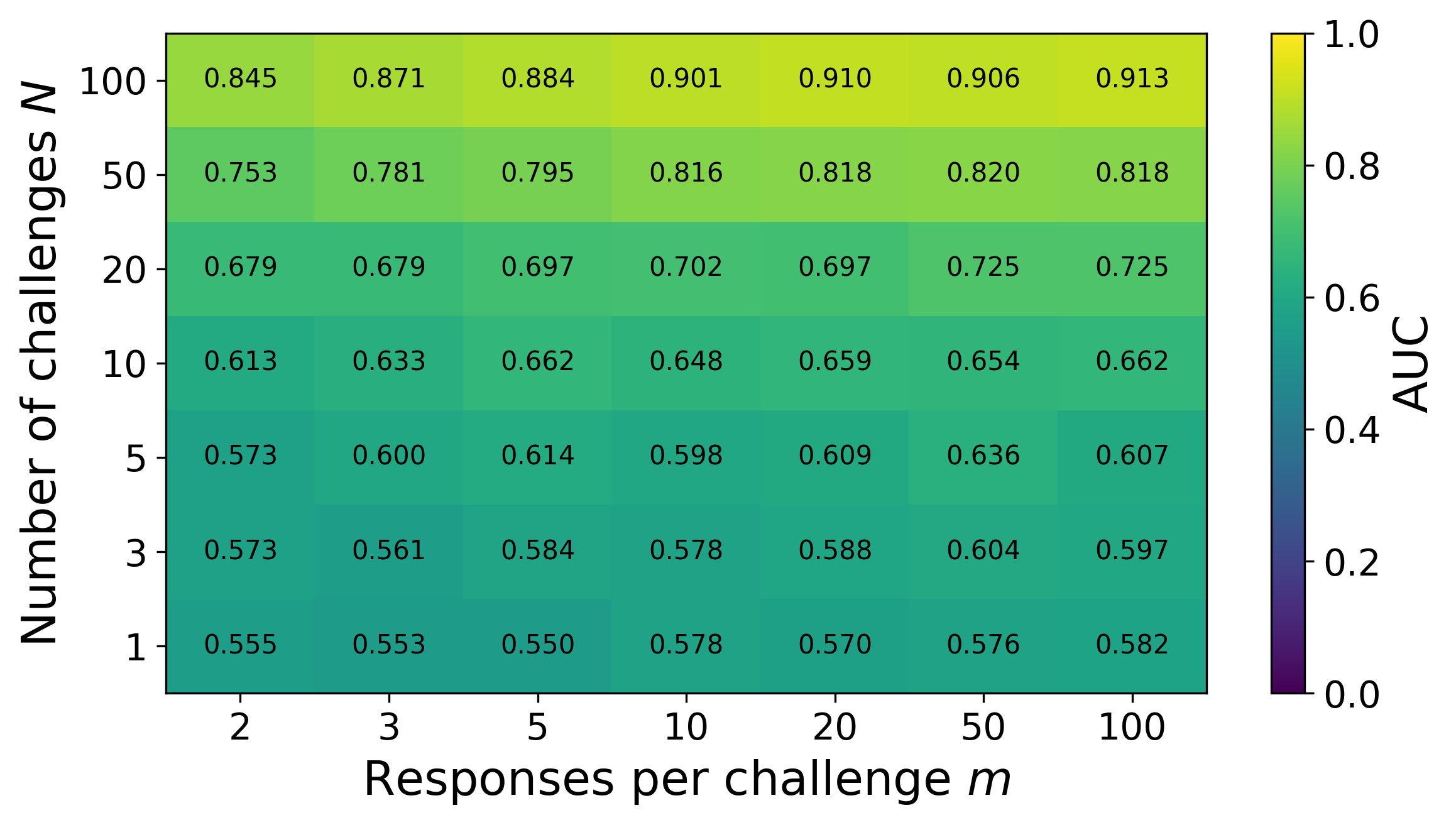}
    \smallskip
    \text{(b) $M^\textrm{MLaaS}$: Llama-2}
\end{minipage}
\caption{
Representative empirical AUC heatmaps for matching--non-matching
separation under different verification budgets $(N,m)$.
}
\label{fig:empirical_auc}
\end{figure}

Figure \ref{fig:empirical_auc} shows representative empirical AUC
heatmaps for Vicuna and Llama-2 under different verification budgets
$(N,m)$.
The empirical AUC exhibits clear model-dependent behavior.
Vicuna, Mistral, and Qwen2.5 achieve strong matching--non-matching
separation with relatively small verification budgets, consistent with
their large probability-control gaps observed in the previous
experiment.
In contrast, Llama-2 and Yi require substantially more accumulated
evidence to achieve comparable separation.

Interestingly, the allocation of the query budget between the number
of independent challenges $N$ and the number of repeated responses
$m$ has a strong effect on verification performance.
For Llama-2 and Yi, when $N$ is small, the AUC remains below $0.7$ even
when $m$ is increased substantially.
In contrast, even with the minimum setting $m=2$, increasing $N$ leads
to a marked improvement in AUC.
Thus, even for comparable total query budgets $Nm$, allocating queries
to a larger number of independent challenges can yield substantially
better separation than repeatedly querying a small number of
challenges.

These results indicate that verification performance is determined not
only by stochastic observation noise within each challenge, but also by
the variation across independently generated challenges.
This relationship between $N$, $m$, and matching--non-matching
separation is theoretically characterized in the following section.

\subsection{Comparison Between Theoretical and Empirical AUC}
\label{subsec:theory_empirical_auc}

We next examine whether the theoretical AUC in
Eq.~\eqref{eq:theoretical_auc} predicts the empirical verification
performance.
For each challenge, $q_i$ is estimated by averaging all stored binary outcomes for that challenge.
The quantities $\mathcal{E}^{X}$ and $A^X,B^X,C^X$ defined in
Eqs.~\eqref{eq:class_error} and \eqref{eq:A_X}--\eqref{eq:C_X}
are then estimated by replacing the latent $q_i^X$ with its empirical
estimate and replacing the corresponding expectations by empirical
averages.

For the matching class $\mathtt{T}$, these averages are taken over the $500$
common challenges evaluated with the matching prover.
For the non-matching class $\mathtt{F}$, the four non-matching prover populations
are combined before estimation, giving a single pooled population of
$4\times500=2000$.
Thus, the $\mathtt{F}$-class parameters are estimated directly from the pooled
population.
Using these estimates in Eq.~\eqref{eq:w_explicit}, the theoretical AUC for
each $(N,m)$ is obtained from Eq.~\eqref{eq:theoretical_auc}.
The theoretical and empirical AUCs are evaluated using the same stored challenge and response data. 
For each Monte Carlo trial and each $(N,m)$, $N$ instances are sampled without replacement from both the matching and pooled non-matching populations. For each selected instance, $m$ observations are sampled without replacement, and the resulting U-statistics are averaged according to Eq.~\eqref{eq:aggregate_score}.
Accordingly, this comparison evaluates how well the Gaussian approximation with the estimated parameters reproduces verification performance on the observed empirical populations. It does not constitute an independent evaluation of prediction accuracy for newly generated challenges or previously unobserved prover populations. The theoretical variance retains the independent-sampling population form in Eq.~\eqref{eq:aggregate_variance}, while the empirical evaluation uses resampling without replacement from the finite stored dataset.
\begin{table}[t]
\centering
\caption{
Parameter Values Computed from the Observed Data\\for the Theoretical
AUC Calculation.
}
\label{tab:estimated_theoretical_parameters}

\renewcommand{\arraystretch}{1.15}

\begin{tabular}{l|c|cccc}
\hline
\textbf{Evaluation model}
&
\textbf{Class}
&
$\widehat{\mathcal{E}}^{X}$
&
$\widehat{A}^X$
&
$\widehat{B}^X$
&
$\widehat{C}^X$
\\
\hline

\multirow{2}{*}{Vicuna}
&
$\mathtt{T}$
&
0.0392
&
0.0083
&
0.0231
&
0.0774
\\
&
$\mathtt{F}$
&
0.2099
&
0.0418
&
0.0421
&
0.0258
\\
\hline

\multirow{2}{*}{Mistral}
&
$\mathtt{T}$
&
0.0541
&
0.0173
&
0.0135
&
0.0671
\\
&
$\mathtt{F}$
&
0.2284
&
0.0462
&
0.0287
&
0.0202
\\
\hline

\multirow{2}{*}{Llama-2}
&
$\mathtt{T}$
&
0.1674
&
0.0431
&
0.0246
&
0.0358
\\
&
$\mathtt{F}$
&
0.2043
&
0.0432
&
0.0342
&
0.0270
\\
\hline

\multirow{2}{*}{Qwen}
&
$\mathtt{T}$
&
0.0420
&
0.0137
&
0.0104
&
0.0685
\\
&
$\mathtt{F}$
&
0.2326
&
0.0477
&
0.0200
&
0.0144
\\
\hline

\multirow{2}{*}{Yi}
&
$\mathtt{T}$
&
0.1411
&
0.0377
&
0.0291
&
0.0463
\\
&
$\mathtt{F}$
&
0.2076
&
0.0424
&
0.0418
&
0.0284
\\
\hline

\end{tabular}
\end{table}
\begin{figure}[t]
\centering
\begin{minipage}[b]{0.49\columnwidth}
    \centering
    \includegraphics[width=\linewidth]{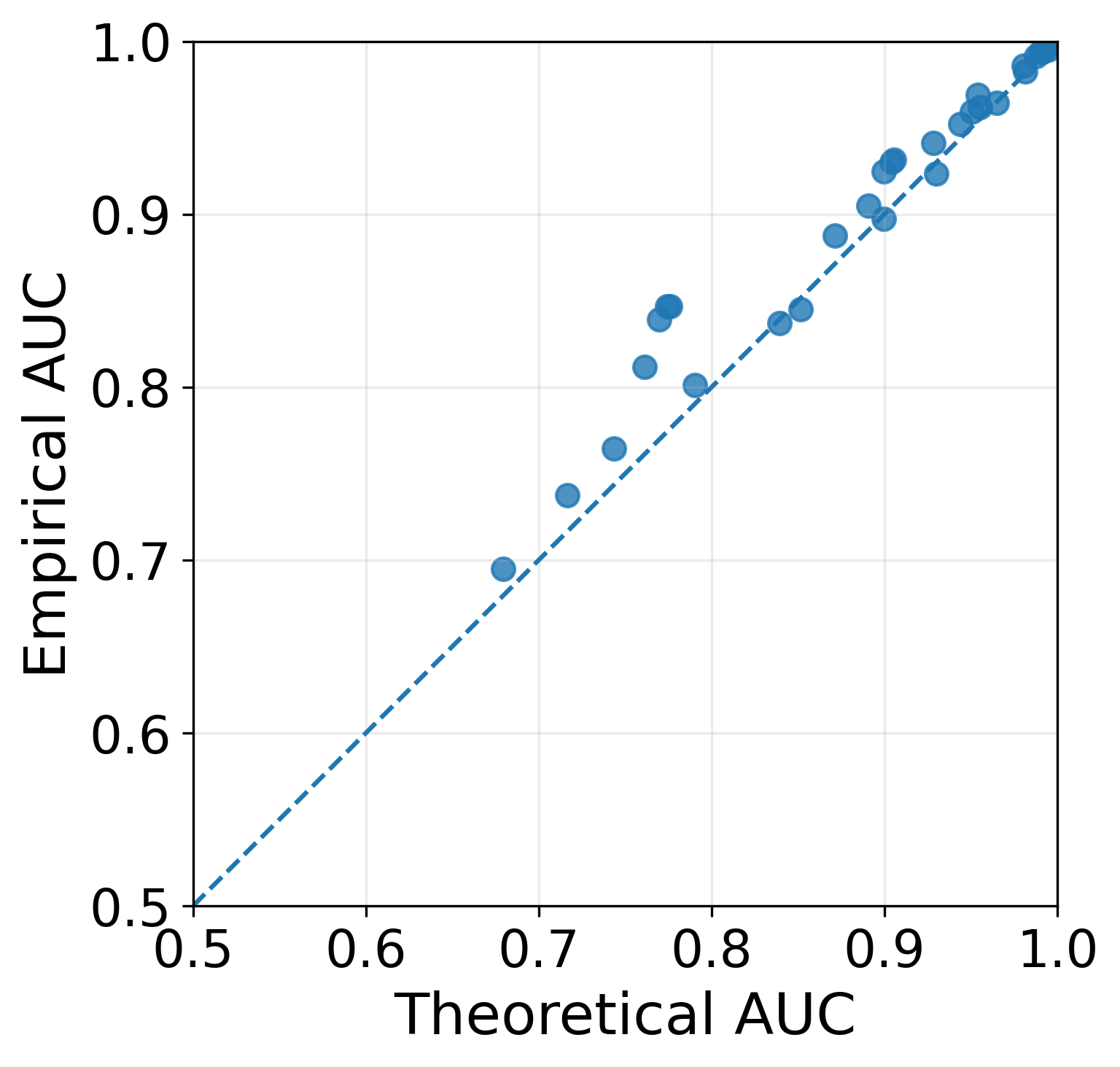}
    \smallskip
    \text{(a) $M^\textrm{MLaaS}$: Vicuna}
\end{minipage}
\begin{minipage}[b]{0.49\columnwidth}
    \centering
    \includegraphics[width=\linewidth]{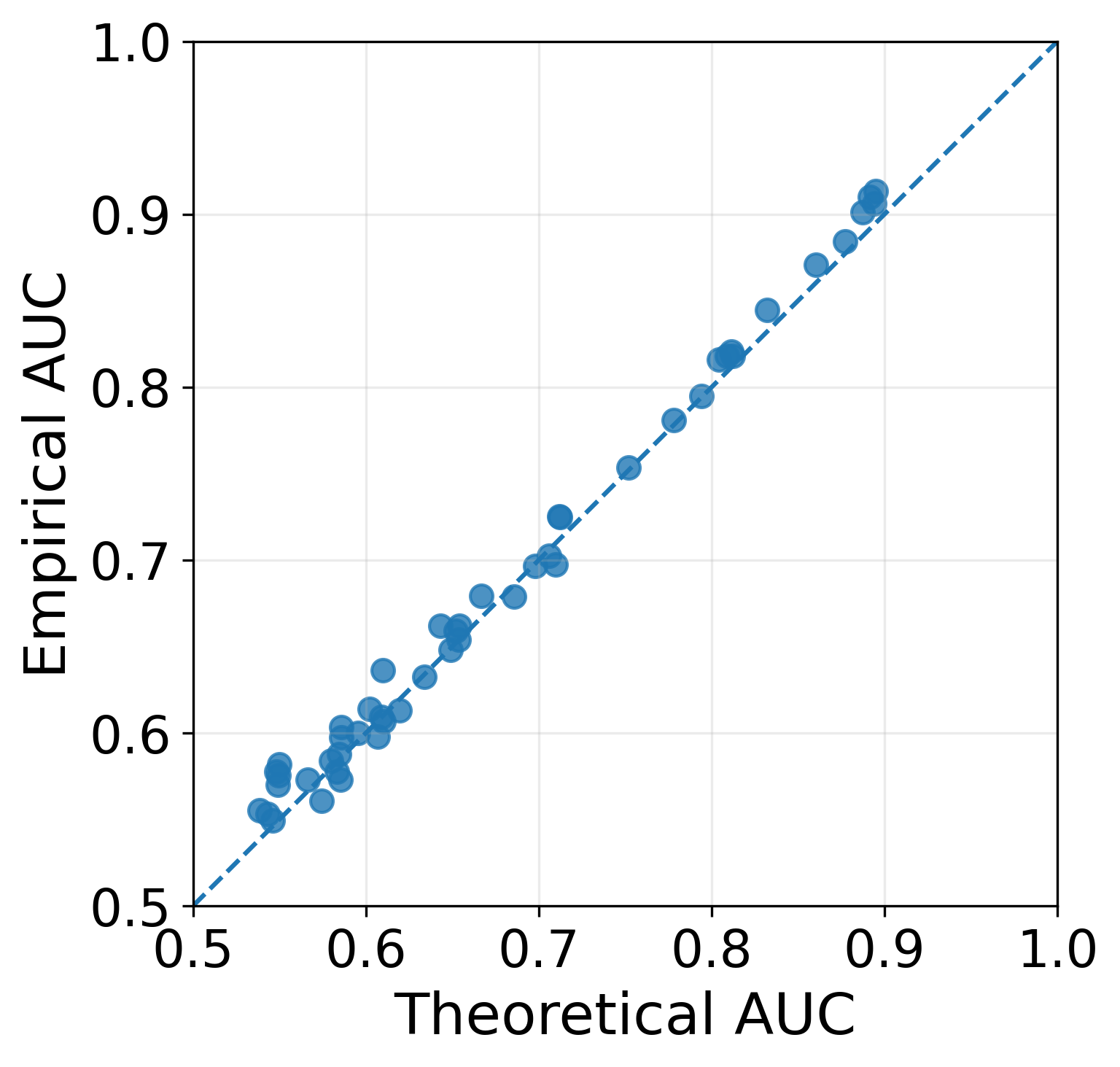}
    \smallskip
    \text{(b) $M^\textrm{MLaaS}$: Llama-2}
\end{minipage}
\caption{
Comparison between theoretical and empirical pooled AUC for different
verification budgets $(N,m)$.
The dashed line indicates perfect agreement.
(a) Vicuna and (b) Llama-2 as
$M^{\mathrm{MLaaS}}$.
}
\label{fig:theoretical_empirical_auc}
\end{figure}

Table~\ref{tab:estimated_theoretical_parameters} lists the parameter
values computed from the observed data and used in the theoretical
AUC calculation.
Figure \ref{fig:theoretical_empirical_auc} compares the theoretical and
empirical pooled AUCs over all tested $(N,m)$ combinations.
The largest deviations from the diagonal are observed for $N=1$,
particularly at large $m$.
At $(N,m)=(1,100)$, the absolute AUC differences are $0.1033$ for
Qwen2.5, $0.1043$ for Mistral, and $0.0707$ for Vicuna.
In these cases, the empirical AUC is higher than the theoretical AUC.
This discrepancy is primarily attributed to the limited accuracy of
the Gaussian approximation for $N=1$, where no aggregation over
multiple challenge-wise U-statistics occurs.
In contrast, aggregating multiple challenges makes the distribution
of $T_{N,m}$ closer to Gaussian.
Indeed, across all evaluated target models and all tested $(N,m)$
combinations with $N>1$, the maximum absolute difference between
the theoretical and empirical AUCs is only $0.028$.
Thus, for the evaluated empirical populations and budget settings with $N>1$, Eq.~\eqref{eq:theoretical_auc} closely reproduces the verification performance obtained by resampling the stored responses.

\subsection{Required Numbers of Challenges and Responses}
\label{subsec:required_budget_experiment}

Finally, we examine whether the theoretical expressions for the required verification budget agree with the empirically observed values.
Using Eqs.~\eqref{eq:nmin_explicit} and \eqref{eq:m_min}, together
with the parameter values estimated in
Table~\ref{tab:estimated_theoretical_parameters}, we calculate the theoretical
minimum verification budgets, $N_{\min}^{\mathrm{theo}}(m,\alpha)$ and
$m_{\min}^{\mathrm{theo}}(N,\alpha)$, for target AUC levels
$\alpha\in\{0.80,0.90,0.95\}$.

For $m_{\min}^{\mathrm{theo}}(N,\alpha)$, the theoretical analysis additionally
determines whether a finite number of responses can achieve the target AUC.
In the present experiment, we evaluate the quantity $R$ defined in
Eq.~\eqref{eq:R_definition} using the estimated values
$\widehat{\mathcal{E}}^{\mathtt{T}}$ and $\widehat{\mathcal{E}}^{\mathtt{F}}$, and denote the
result by
\begin{equation}
\widehat{R}(N,\alpha)
=
\frac{
N
\left(
\widehat{\mathcal{E}}^{\mathtt{F}}
-
\widehat{\mathcal{E}}^{\mathtt{T}}
\right)^2
}{
\left(\Phi^{-1}(\alpha)\right)^2
}.
\label{eq:experimental_R_definition}
\end{equation}
Similarly, we define
$\widehat{A}=\widehat{A}^{\mathtt{T}}+\widehat{A}^{\mathtt{F}}$.
A finite $m_{\min}^{\mathrm{theo}}(N,\alpha)$ exists if and only if
\begin{equation}
\widehat{A}
<
\widehat{R}(N,\alpha).
    \label{eq:experimental_m_feasibility}
\end{equation}
When this condition is not satisfied,
no finite $m$ satisfies the target-AUC condition under the theoretical
approximation; such cases are denoted by $\times$.

For comparison, we directly obtain the corresponding minimum budgets
from the empirical pooled AUC.
For a fixed number of responses $m$, we define
$N_{\min}^{\mathrm{emp}}(m,\alpha)$ as the smallest integer
$N=1,\ldots,100$ for which the empirical AUC is strictly greater than
the target $\alpha$.
Similarly, for a fixed number of challenges $N$, we define
$m_{\min}^{\mathrm{emp}}(N,\alpha)$ as the smallest integer
$m=2,\ldots,100$ for which the empirical AUC is strictly greater than
$\alpha$.
The same pooled matching--non-matching score distributions and sampling
procedure used in the preceding empirical AUC experiment are used for
this search.
If the target AUC is not exceeded within the corresponding search
range, the empirical result is denoted by $\mathrm{NR}$ (not reached).

% \begin{figure}[t]
% \centering
% \begin{minipage}[b]{0.49\linewidth}
%     \centering
%     \includegraphics[width=\linewidth]
%     {figs/ex_minN_minm/mini_version/vicuna_min_m_vs_N_multiAUC_with_empirical.png}
% \end{minipage}
% \hfill
% \begin{minipage}[b]{0.49\linewidth}
%     \centering
%     \includegraphics[width=\linewidth]
%     {figs/ex_minN_minm/mini_version/vicuna_min_N_vs_m_multiAUC_with_empirical.png}
% \end{minipage}

% \smallskip
% \text{(a) Vicuna}

% \vspace{0.8em}

% \begin{minipage}[b]{0.49\linewidth}
%     \centering
%     \includegraphics[width=\linewidth]
%     {figs/ex_minN_minm/mini_version/llama_min_m_vs_N_multiAUC_with_empirical.png}
% \end{minipage}
% \hfill
% \begin{minipage}[b]{0.49\linewidth}
%     \centering
%     \includegraphics[width=\linewidth]
%     {figs/ex_minN_minm/mini_version/llama_min_N_vs_m_multiAUC_with_empirical.png}
% \end{minipage}

% \smallskip
% \text{(b) Llama-2}

% \caption{
% Representative comparisons of empirical and theoretical minimum
% verification budgets for target AUC levels
% $\alpha\in\{0.80,0.90,0.95\}$.
% For each $M^{\mathrm{MLaaS}}$, the left panel shows
% $m_{\min}(N,\alpha)$ as a function of $N$, while the right panel shows
% $N_{\min}(m,\alpha)$ as a function of $m$.
% Solid curves indicate the theoretical estimates and cross markers
% indicate the corresponding empirical minima.
% No empirical marker is shown for conditions denoted by $\mathrm{NR}$.
% (a) Vicuna and (b) Llama-2.
% }
% \label{fig:min_N_min_m}
% \end{figure}

\begin{table*}[t]
\centering
\caption{
Empirical and theoretical minimum verification budgets:
$N_{\min}$ for fixed $m$ and $m_{\min}$ for fixed $N$.
$\mathrm{NR}$ denotes that the target AUC was not reached within the search range, whereas $\times$ denotes that no finite $m_{\min}^{\mathrm{theo}}$ exists.
}
\label{tab:min_N_m_comparison}
\scriptsize
\renewcommand{\arraystretch}{1.15}
\setlength{\tabcolsep}{2.0pt}
\begin{tabular}{c c | c c c c c c c | c c c c c c c}
\hline
\multirow{3}{*}{$M^{\mathrm{MLaaS}}$}
&
\multirow{3}{*}{$\alpha$}
&
\multicolumn{7}{c|}{$N_{\min}^{\mathrm{emp}}(m,\alpha) \,/\, N_{\min}^{\mathrm{theo}}(m,\alpha)$}
&
\multicolumn{7}{c}{$m_{\min}^{\mathrm{emp}}(N,\alpha) \,/\, m_{\min}^{\mathrm{theo}}(N,\alpha)$}
\\
\cline{3-16}
&
&
$m=2$
&
$m=3$
&
$m=5$
&
$m=10$
&
$m=20$
&
$m=50$
&
$m=100$
&
$N=1$
&
$N=3$
&
$N=5$
&
$N=10$
&
$N=20$
&
$N=50$
&
$N=100$
\\
\hline

\multirow{3}{*}{Vicuna}
& 0.80
& 3 / 4 & 2 / 3 & 2 / 2 & 1 / 2 & 1 / 2 & 1 / 2 & 1 / 2
& 8 / $\times$ & 2 / 3 & 2 / 2 & 2 / 2 & 2 / 2 & 2 / 2 & 2 / 2 \\
& 0.90
& 8 / 8 & 6 / 6 & 4 / 4 & 3 / 4 & 3 / 4 & 3 / 3 & 3 / 3
& $\mathrm{NR}$ / $\times$ & 8 / 22 & 4 / 4 & 2 / 2 & 2 / 2 & 2 / 2 & 2 / 2 \\
& 0.95
& 13 / 13 & 8 / 9 & 6 / 7 & 5 / 6 & 5 / 5 & 4 / 5 & 5 / 5
& $\mathrm{NR}$ / $\times$ & $\mathrm{NR}$ / $\times$ & 7 / 19 & 3 / 3 & 2 / 2 & 2 / 2 & 2 / 2 \\
\hline

\multirow{3}{*}{Mistral}
& 0.80
& 3 / 3 & 2 / 3 & 2 / 2 & 1 / 2 & 1 / 2 & 1 / 2 & 1 / 2
& 8 / $\times$ & 2 / 2 & 2 / 2 & 2 / 2 & 2 / 2 & 2 / 2 & 2 / 2 \\
& 0.90
& 7 / 7 & 5 / 5 & 4 / 5 & 4 / 4 & 3 / 4 & 3 / 4 & 3 / 4
& $\mathrm{NR}$ / $\times$ & 14 / $\times$ & 3 / 3 & 2 / 2 & 2 / 2 & 2 / 2 & 2 / 2 \\
& 0.95
& 12 / 12 & 8 / 9 & 7 / 7 & 6 / 7 & 6 / 6 & 6 / 6 & 6 / 6
& $\mathrm{NR}$ / $\times$ & $\mathrm{NR}$ / $\times$ & 76 / $\times$ & 3 / 3 & 2 / 2 & 2 / 2 & 2 / 2 \\
\hline

\multirow{3}{*}{Llama-2}
& 0.80
& 68 / 77 & 55 / 61 & 47 / 53 & 47 / 49 & 45 / 47 & 41 / 46 & 41 / 46
& $\mathrm{NR}$ / $\times$ & $\mathrm{NR}$ / $\times$ & $\mathrm{NR}$ / $\times$ & $\mathrm{NR}$ / $\times$ & $\mathrm{NR}$ / $\times$ & 6 / 7 & 2 / 2 \\
& 0.90
& $\mathrm{NR}$ / 178 & $\mathrm{NR}$ / 141 & $\mathrm{NR}$ / 122 & 99 / 112 & 92 / 108 & 90 / 106 & 90 / 105
& $\mathrm{NR}$ / $\times$ & $\mathrm{NR}$ / $\times$ & $\mathrm{NR}$ / $\times$ & $\mathrm{NR}$ / $\times$ & $\mathrm{NR}$ / $\times$ & $\mathrm{NR}$ / $\times$ & 6 / $\times$ \\
& 0.95
& $\mathrm{NR}$ / 292 & $\mathrm{NR}$ / 231 & $\mathrm{NR}$ / 201 & $\mathrm{NR}$ / 185 & $\mathrm{NR}$ / 178 & $\mathrm{NR}$ / 174 & $\mathrm{NR}$ / 173
& $\mathrm{NR}$ / $\times$ & $\mathrm{NR}$ / $\times$ & $\mathrm{NR}$ / $\times$ & $\mathrm{NR}$ / $\times$ & $\mathrm{NR}$ / $\times$ & $\mathrm{NR}$ / $\times$ & $\mathrm{NR}$ / $\times$ \\
\hline

\multirow{3}{*}{Qwen}
& 0.80
& 3 / 3 & 2 / 2 & 1 / 2 & 1 / 2 & 1 / 2 & 1 / 2 & 1 / 2
& 5 / $\times$ & 2 / 2 & 2 / 2 & 2 / 2 & 2 / 2 & 2 / 2 & 2 / 2 \\
& 0.90
& 6 / 6 & 4 / 4 & 4 / 4 & 3 / 3 & 2 / 3 & 2 / 3 & 2 / 3
& $\mathrm{NR}$ / $\times$ & 6 / 9 & 3 / 3 & 2 / 2 & 2 / 2 & 2 / 2 & 2 / 2 \\
& 0.95
& 9 / 9 & 6 / 7 & 5 / 6 & 5 / 5 & 5 / 5 & 5 / 5 & 5 / 5
& $\mathrm{NR}$ / $\times$ & $\mathrm{NR}$ / $\times$ & 5 / 8 & 2 / 2 & 2 / 2 & 2 / 2 & 2 / 2 \\
\hline

\multirow{3}{*}{Yi}
& 0.80
& 23 / 25 & 19 / 20 & 14 / 16 & 14 / 15 & 13 / 14 & 14 / 14 & 12 / 13
& $\mathrm{NR}$ / $\times$ & $\mathrm{NR}$ / $\times$ & $\mathrm{NR}$ / $\times$ & $\mathrm{NR}$ / $\times$ & 3 / 3 & 2 / 2 & 2 / 2 \\
& 0.90
& 53 / 57 & 42 / 44 & 36 / 37 & 31 / 33 & 30 / 32 & 30 / 31 & 28 / 30
& $\mathrm{NR}$ / $\times$ & $\mathrm{NR}$ / $\times$ & $\mathrm{NR}$ / $\times$ & $\mathrm{NR}$ / $\times$ & $\mathrm{NR}$ / $\times$ & 3 / 3 & 2 / 2 \\
& 0.95
& 85 / 94 & 68 / 72 & 53 / 60 & 50 / 54 & 47 / 52 & 43 / 50 & 45 / 50
& $\mathrm{NR}$ / $\times$ & $\mathrm{NR}$ / $\times$ & $\mathrm{NR}$ / $\times$ & $\mathrm{NR}$ / $\times$ & $\mathrm{NR}$ / $\times$ & 10 / 41 & 2 / 2 \\
\hline
\end{tabular}
\end{table*}

To quantify the numerical agreement when both minimum budgets are
finite, we use the mean absolute error (MAE) as the primary metric.
Because the magnitude of the required budget differs substantially
across target models, we additionally use the mean relative absolute
error (MRAE) to measure the error relative to the scale of the
theoretical prediction.
For $N_{\min}$, these metrics are defined as
\begin{equation}
    \mathrm{MAE}_{{N_{\mathrm{min}}}}
=
\frac{1}{K}
\sum_{i=1}^{K}
\left|
N_{\min,i}^{\mathrm{emp}}
-
N_{\min,i}^{\mathrm{theo}}
\right|,
\label{eq:budget_mae}
\end{equation}
\begin{equation}
    \mathrm{MRAE}_{{N_{\mathrm{min}}}}
=
\frac{100}{K}
\sum_{i=1}^{K}
\frac{
\left|
N_{\min,i}^{\mathrm{emp}}
-
N_{\min,i}^{\mathrm{theo}}
\right|
}{
N_{\min,i}^{\mathrm{theo}}
}.
\label{eq:budget_mrae}
\end{equation}
Here, $K$ denotes the number of conditions for which both
$N_{\min}^{\mathrm{emp}}$ and $N_{\min}^{\mathrm{theo}}$ are finite.
The corresponding metrics for $m_{\min}$ are defined analogously by
replacing $N_{\min}$ with $m_{\min}$.

\begin{table}[t]
\centering
\caption{
Errors between empirical and theoretical minimum verification budgets.
}
\label{tab:min_budget_error}
\scriptsize
\renewcommand{\arraystretch}{1.12}
\setlength{\tabcolsep}{2.3pt}
\begin{tabular}{c | c c c | c c c}
\hline
\multirow{2}{*}{$M^{\mathrm{MLaaS}}$}
&
\multicolumn{3}{c|}{$N_{\min}$}
&
\multicolumn{3}{c}{$m_{\min}$}
\\
\cline{2-7}
&
$K$
&
$\mathrm{MAE}_{{N_{\mathrm{min}}}}$
&
$\mathrm{MRAE}_{{N_{\mathrm{min}}}}$ (\%)
&
$K$
&
$\mathrm{MAE}_{{m_{\mathrm{min}}}}$
&
$\mathrm{MRAE}_{{m_{\mathrm{min}}}}$ (\%)
\\
\hline
Vicuna
& 21 & 0.571 & 17.64
& 17 & 1.588 & 9.42
\\
Mistral
& 21 & 0.524 & 16.84
& 15 & 0.000 & 0.00
\\
Llama-2
& 11 & 8.636 & 10.79
& 2 & 0.500 & 7.14
\\
Qwen
& 21 & 0.476 & 18.14
& 17 & 0.353 & 4.17
\\
Yi
& 21 & 3.000 & 7.20
& 7 & 4.429 & 10.80
\\
\hline
\end{tabular}
\end{table}

Table~\ref{tab:min_N_m_comparison} summarizes the empirical and theoretical minimum budgets for all five target models. We discuss separately the cases in which both minima are finite and those in which at least one minimum is unavailable.

First, when both the empirical and theoretical minimum budgets are
finite, their numerical agreement is evaluated using the MAE and MRAE.
Table~\ref{tab:min_budget_error} summarizes the results.
For Vicuna, Mistral, and Qwen, the theoretical $N_{\min}$ estimates differ from the empirical minima by less than one challenge on average.
For Llama-2 and Yi, the larger absolute errors mainly reflect the larger scale of the required $N_{\min}$, while the relative deviations remain modest.
A similar agreement is observed for $m_{\min}$, with the theoretical estimates remaining close to the empirical minima across the finite cases.
Overall, the finite-pair results show good agreement between the theoretical and empirical minimum verification budgets.

Second, $\mathrm{NR}$ indicates only that the target AUC was not reached within the empirical search range. For Llama-2, several theoretical $N_{\min}$ values exceed $100$, which is consistent with the corresponding empirical $\mathrm{NR}$ results under the search limit $N\leq100$.

In contrast, a theoretical $\times$ for $m^\mathrm{theo}_{\min}$ indicates that
Eq.~\eqref{eq:experimental_m_feasibility} is not satisfied for the
estimated parameters.
In this case, the theoretical AUC remains below the target even as
$m\rightarrow\infty$, because increasing the number of responses
eliminates the finite-response variance terms but not the intrinsic
challenge-to-challenge variance represented by $\widehat{A}$.
When an empirical $\mathrm{NR}$ is paired with a theoretical
$\times$, the two results are therefore qualitatively consistent:
the theory predicts that increasing $m$ alone cannot achieve the
target AUC, while the empirical search does not achieve it within
$m\leq100$.
The empirical result alone, however, does not establish infeasibility
beyond the investigated range.

Several conditions instead yield a finite $m^\mathrm{emp}_{\min}$ while
$m^\mathrm{theo}_{\min}$ is $\times$.
These cases correspond to disagreement in whether the target-AUC
boundary is crossed, rather than simply to a large numerical error in
the predicted budget.
The theoretical feasibility decision is based on the Gaussian AUC
approximation and the estimated population parameters.
Consequently, when the limiting AUC is close to the target $\alpha$,
a relatively small approximation error can change the binary
finite/infeasible classification.
This effect is particularly relevant for small $N$, for which the
Gaussian approximation of the aggregate score is less accurate, as
also observed in the preceding comparison between theoretical and
empirical AUC.
Thus, the finite-empirical/theoretical-$\times$ cases are naturally
interpreted as boundary cases in which the feasibility classification
is more sensitive to approximation error than the numerical budget
estimation in the finite-pair regime.

%%%%%%%%%%%%%%%%%%%%%%%%%%%%%%%%%%%%%%%%

\section{Concluding Remarks and Future Work}
\label{sec:conclusion}
This work characterized statistical separability in TP-CRIV for probabilistic AI models.
We treated stochastic response behavior as an explicit control target
rather than merely as observation noise and introduced a U-statistic
for unbiasedly observing the resulting challenge-wise probabilistic
discrepancy.
By analyzing the verification score, we derived a
theoretical approximation of the AUC and further connected the
challenge-wise probabilistic behavior to the verification budget
required to achieve a prescribed detection performance.
Thus, the main contribution of this work is a statistical
characterization that relates model-dependent probabilistic behavior
to verification-level separability and the required verification
budget.

We instantiated the proposed characterization for LLMs
using open-ended challenges.
Experiments demonstrated matching--non-matching
separation and showed that the theoretical AUC closely follows the
empirical verification performance.
The theoretically estimated minimum verification budgets were also
consistent with the corresponding empirical minima over the finite
regime.
These results demonstrate that probabilistic challenge--response
behavior, inferred from stochastic observations, can be quantitatively related to verification performance
and used to estimate the amount of evidence required for verification.

This study establishes a theoretical characterization of statistical
separability as a starting point for investigating probabilistic
challenge--response verification under broader conditions.
Future work will extend the empirical scope to broader prover models,
witness-generation strategies, and challenge designs, with the aim of
characterizing the challenge-wise probabilistic behaviors that arise
under different verification conditions.
The theoretical characterization developed in this work can then be
used to translate these behaviors into expected
matching--non-matching separability and required verification budgets.
Such evaluations will also help identify challenge designs and prover
conditions that yield reliable separation in practice.

%%%%%%%%%%%%%%%%%%%%%%%%%%%%%%%%%%%%%%%%

\bibliographystyle{ieeetr}
\bibliography{refs_short}

\appendix
\section{LLM Experimental Settings}
\label{app:llm_settings}

This appendix provides the experimental settings shared across all
evaluated LLMs and the prompt templates used for each model.

\subsection{Common Settings}
\label{app:common_settings}

Table~\ref{tab:llm_settings} summarizes the settings used for all five
models.
The same inference settings are used for both prover-side and
MLaaS-side evaluations, and the same suffix-generation settings
are used by all provers.

\begin{table}[t]
\centering
\caption{Common experimental settings for all evaluated LLMs.}
\label{tab:llm_settings}
\begin{tabular}{ll}
\hline
\textbf{Parameter} & \textbf{Value} \\
\hline
\multicolumn{2}{l}{\textit{Inference settings}} \\
Temperature & $0.8$ \\
Top-$p$ & $0.95$ \\
Top-$k$ & $50$ \\
Maximum new tokens & $32$ \\
Quantization & 4-bit NF4 \\
Computation dtype & FP16 \\
\hline
\multicolumn{2}{l}{\textit{Suffix-generation settings}} \\
GCG suffix length & $1$ token \\
GCG iterations & $12$ \\
GCG top-$k$ candidates & $512$ \\
Directional candidates & $128$ \\
Visible suffix limit $N_s$ & $5$ characters \\
Minimum prohibited fragment length & $3$ characters \\
Empirical shortlist size & $50$ \\
First-stage samples $K_1$ & $20$ \\
First-stage retained candidates & $20$ \\
Additional samples $K_2$ & $80$ \\
\hline
\end{tabular}
\end{table}

\subsection{Model-Specific Prompt Templates}
\label{app:prompt_templates}

The serialization function $\mathcal{T}_M$ follows a fixed template
associated with the model receiving the query.
Each model uses the same template in both its prover and MLaaS roles.

For all models, we use the same system instruction $h$, inserted
into the \texttt{\{system\_prompt\}} field of each model's template.
Here, \texttt{\{base\_prompt\}} and \texttt{\{suffix\}} denote
$x_i$ and $s_i$, respectively.
The exact prompt formats are given below.

\subsubsection{Mistral-7B-Instruct-v0.2}
\begin{verbatim}
[INST] {system_prompt}
{base_prompt}{suffix} [/INST]
\end{verbatim}

\subsubsection{Vicuna-7B-v1.5}
\begin{verbatim}
{system_prompt}
USER: {base_prompt}{suffix} ASSISTANT:
\end{verbatim}

\subsubsection{Llama-2-7B-chat-hf}
\begin{verbatim}
[INST] <<SYS>>
{system_prompt}
<</SYS>>
{base_prompt}{suffix} [/INST]
\end{verbatim}

\subsubsection{Qwen2.5-7B-Instruct and Yi-6B-Chat}
\begin{verbatim}
<|im_start|>system
{system_prompt}<|im_end|>
<|im_start|>user
{base_prompt}{suffix}<|im_end|>
<|im_start|>assistant
\end{verbatim}

\vfill

\end{document}